\documentclass[12pt]{article}

\usepackage{xspace}
\usepackage{epsfig}

\newcommand{\nc}{\newcommand}

\nc{\beq}{\begin{equation}}
\nc{\eeq}{\end{equation}}
\nc{\beqa}{\begin{eqnarray}}
\nc{\eeqa}{\end{eqnarray}}
\nc{\bea}{\begin{eqnarray}}
\nc{\eea}{\end{eqnarray}}
\nc{\ra}{\rightarrow}
\nc{\lsim}{\begin{array}{c}\,\sim\vspace{-21pt}\\< \end{array}}
\nc{\gsim}{\begin{array}{c}\sim\vspace{-21pt}\\> \end{array}}

\nc{\LL}{L}
\nc{\vv}{\tilde{v}}
\nc{\GG}{\tilde{G}}

\title{
\vspace*{-1.3cm}
\begin{flushright}
\normalsize{
ANL-HEP-PR-03-029\\
EFI-03-18\\
FERMILAB-PUB-03/066-T
  }
\end{flushright}
\vspace{1.5cm}
\Large
\textbf{Precision Electroweak Data and Unification of \\
Couplings in Warped Extra Dimensions}\vspace*{1.0cm}
\author{\large\textbf{Marcela Carena$^a$}, \textbf{Antonio Delgado$^b$},
\textbf{Eduardo Pont\'{o}n$^a$},\\[0.3cm]
\textbf{Tim M.P. Tait$^a$},
and 
\textbf{C.E.M.~Wagner$^{c,d}$}\\ \\[0.5cm]
$^a$\normalsize\emph{Fermi National Accelerator Laboratory,
P.O. Box 500, Batavia, IL 60510, USA} \\
$^b$\normalsize\emph{Department of Physics and Astronomy, Johns Hopkins University,}\\ \normalsize\emph{3400 North Charles St., Baltimore, MD 21218, USA} \\
$^c$\normalsize\emph{HEP Division, Argonne National Laboratory,
9700 Cass Ave.,
Argonne, IL 60439, USA} \\
$^d$\normalsize\emph{Enrico Fermi Institute, Univ. of Chicago, 5640
Ellis Ave., Chicago, IL 60637, USA}}}

\begin{document}
\setcounter{page}{0}
\maketitle
\begin{abstract}
Warped extra dimensions allow a novel way of solving the hierarchy
problem, with all fundamental mass parameters of the theory naturally
of the order of the Planck scale.  The observable value of the Higgs
vacuum expectation value is red-shifted, due to the localization of
the Higgs field in the extra dimension.  It has been recently observed
that, when the gauge fields propagate in the bulk, unification of the
gauge couplings may be achieved.  Moreover, the propagation of
fermions in the bulk allows for a simple solution to potentially
dangerous proton decay problems.  However, bulk gauge fields and
fermions pose a phenomenological challenge, since they tend to induce
large corrections to the precision electroweak observables.  In this
article, we study in detail the effect of gauge and fermion fields
propagating in the bulk in the presence of gauge brane kinetic terms
compatible with gauge coupling unification, and we present ways of
obtaining a consistent description of experimental data, while
allowing values of the first Kaluza Klein mode masses of the order of
a few TeV.
\end{abstract}

\thispagestyle{empty}
\newpage

\setcounter{page}{1}

\baselineskip18pt

\section{Introduction}

Large extra dimensions have been introduced as a way of solving the
hierarchy problem, related to the hierarchy between the Planck and the
weak scales.  In flat extra dimensions, this can be achieved by
assuming that the Standard Model (SM) fields are localized on a four
dimensional brane, while gravity propagates in the extra dimensional
bulk \cite{Arkani-Hamed:1998rs}.  Under such conditions, the
fundamental Planck scale may be of order of the weak scale.  The
weakness of the gravitational interactions is explained by the large
volume suppression of the zero mode graviton interactions with the
localized SM fields.  For this mechanism to work, the size of the
extra dimensions must be several orders of magnitude larger than the
fundamental scale of the model.

An interesting alternative is achieved in the case of warped extra
dimensions proposed by Randall and Sundrum (RS) \cite{Randall:1999ee}. 
In this case, one can obtain a solution to the hierarchy problem by
assuming that the gravity field propagates in the bulk of a slice of
5-dimensional anti de Sitter space (AdS$_5$) bounded by 4-dimensional
``branes", while the Higgs field is localized on the brane where the
warp factor is small (the TeV or IR brane).  In this case, all mass
parameters, the fundamental Planck scale, the curvature and the size
of the extra dimensions, as well as the Higgs vacuum expectation value
(VEV), are of the same order.  Due to the nontrivial warp factor,
however, the observable Higgs VEV is red-shifted to values much
smaller than the fundamental Planck scale.

In the minimal model of warped extra dimensions only gravity
propagates in the bulk, while the SM gauge and fermion fields are
localized on the same brane as the Higgs field.  The solution to the
hierarchy problem, however, demands only the localization of the Higgs
field, and there is strong motivation to consider the propagation of
the gauge fields in the bulk of the extra dimension.  In particular,
at scales below the AdS curvature $k$, the gauge couplings evolve
logarithmically \cite{gaugerunning,Goldberger:2002cz}, and unification
of couplings in these scenarios may be naturally achieved whenever the
gauge fields propagate in the bulk \cite{Agashe:2002pr}
(for a related discussion in a supersymmetric context, see 
\cite{Goldberger:2002pc}).  Bulk gauge
fields are, however, phenomenologically challenging since they tend to
induce large corrections to the precision electroweak observables. 
These corrections are induced by a combination of two effects: for
one, the presence of a localized Higgs VEV induces a repulsion of the
zero mode of the gauge fields from the brane location, implying a
modification of the weak gauge boson couplings to the brane fields and
of the relation between the tree-level weak gauge boson masses and the
Higgs VEV. Since these Higgs localization effects are absent for the
photon field, and are different for the $W$ and the $Z$ gauge bosons,
this induces important tree-level corrections to the precision
electroweak observables \cite{Huber:2001gw,Csaki:2002gy}. 
Furthermore, the Kaluza-Klein (KK) modes of the gauge fields couple to
the brane fields with a coupling which is $\sqrt{2 k L}$ times larger
than the zero mode gauge coupling, where $L$ is the proper size of the
extra dimension.  The dimensionless factor $k\,L$ has to be about 30
so that the Higgs VEV is red-shifted from the Planck to the TeV scale. 
Therefore, the KK modes couple strongly to the brane quark and lepton
fields and can induce large effects on the precision electroweak
observables.

The above effects imply a lower bound on the mass of the lightest KK
mode of the gauge fields of the order of $27$ TeV \cite{Csaki:2002gy}. 
Such a large bound excludes the possibility of direct detection of
these KK modes at the Tevatron and the LHC. It also reintroduces to
some extent the hierarchy problem that the RS scenario intended to
solve.  However, in Refs.~\cite{Davoudiasl:2002ua,Carena:2002dz} it
was shown that these bounds may be relaxed by considering the possible
effect of brane localized kinetic terms for the gauge fields
\cite{Carena:2002me}.  Large brane kinetic terms have the effect of
enhancing the four dimensional properties of the gauge fields as seen
by brane observers.  In particular, the presence of gauge kinetic
terms on the IR brane, where the Higgs and fermion fields are
localized, translates into a reduction of the Higgs localization
effects, as well as of the effective coupling of the heavy KK modes
with respect to the one of the zero modes.  In addition, the KK masses
depend strongly on the IR brane kinetic term and can decrease by up to
a factor of ten when these terms are large (and positive).  As a
result, the bound on the mass of the lightest gauge KK mode mass may
be relaxed down to a few TeV. Moreover, if the heavy flavor Z-pole
observables are excluded from the data, it was found that, for certain
values of the weak gauge boson KK masses and of the local gauge
kinetic terms, the fit to the precision electroweak observables
improves with respect to the one obtained within the Standard Model
with no extra dimensions.

In this work we are interested in considering the scenario of warped
extra dimensions in the light of grand unification.  The above
described scenario, with the lightest KK modes in the few TeV range,
requires large brane kinetic terms on the IR brane.  However, we will
find that it is necessary to assume that the GUT symmetry is broken on
the IR brane, which implies that the local kinetic terms need not be
unified.  Indeed, generically the brane kinetic terms include bare
contributions, which encode the physics at or above the cutoff scale,
as well as radiative contributions that are calculable within the
effective 5-dimensional theory.  On the IR brane, the latter effects
are small since the local cutoff is not much above the TeV scale and
there is no room for a large logarithmic enhancement.  Hence, a large
IR brane kinetic term would have to arise from the unknown UV physics
and it becomes necessary to assume that this physics respects the GUT
symmetry.  The situation is different for the UV brane kinetic terms
since there the local cutoff is of the order of the Planck scale and
the loop contributions are logarithmically enhanced at low energies. 
In fact, the large GUT violating logarithms that are found in the
calculation of the low-energy gauge couplings \cite{gaugerunning} can
be understood as arising from the RG flow of the UV brane kinetic
terms, from the Planck to the TeV scale \cite{Goldberger:2002cz}.  On
the other hand, most models of warped unification seem to require
orbifold breaking of the gauge symmetry.  Thus, to retain the
predictivity of these grand unified scenarios (e.g. for the Weinberg
angle), it is necessary to argue that the bare contributions to the
brane kinetic terms are small.  One such argument is furnished by
naive dimensional analysis (NDA) \cite{Manohar:1983md}, which assumes
strong coupling at the cut-off scale and demands brane terms to be
small.  Nevertheless, it should be kept in mind that the IR kinetic
terms could be larger than what naive considerations would suggest,
and have important effects for the spectrum of lowest lying KK modes.

A further complication comes from the fact that the RS unified model
has TeV mass KK gauge bosons which can mediate proton decay.  In fact,
since the effective cut-off on the IR brane is TeV, even
non-renormalizable operators can lead to unacceptably short proton
lifetimes.  These problems may be alleviated by assuming that the
quarks and leptons come from different multiplets, and their partners
are projected out by particular boundary conditions.  Clearly, this
idea can only work if fermions also propagate in the bulk.  In
principle, fermions propagating in the bulk tend to improve the two
phenomenological challenges associated with bulk gauge fields, since
these are mostly related to the localization of the fermion fields in
the same brane as the Higgs field.  However, it has been suggested
that bulk fermions do not lead to a dramatic relaxation of these
constraints \cite{Burdman:2002gr}.  Moreover, the bulk fermions induce
new contributions to the precision electroweak
observables~\cite{Hewett:2002fe}, which can only be suppressed if at
least some of the third generation fermions are localized close to the
infrared brane.

In this article, we shall analyze all these problems in detail and we
shall demonstrate that KK mode masses of the order of several TeV can
be achieved by assuming either a non-vanishing IR gauge kinetic term
for the gauge fields, or a large Higgs mass, together with a
particular geography of the fermion fields in the extra dimensional
bulk.  In section 2, we describe the model and introduce the question
of unification.  In section 3 we present a formalism to study the
experimental implications in a simple and straightforward way.  In
section 4, we analyze the novel effects to precision electroweak
observables due to the Higgs localization effects and the contribution
of the KK tower of gauge bosons and fermions in the bulk.  In section
5, we discuss the fit to the precision electroweak data.  We reserve
section 6 for our conclusions.

\section{Grand Unification}

In this section we review the important features of grand unification
in RS. While we envision the non-supersymmetric model of
Ref.~\cite{Agashe:2002pr}, our framework is quite general, defined by
the low energy theory, and thus we expect our conclusions to hold more
generally for the RS unified framework.  The background metric is
defined by the line element,
\beq
\label{lineelement}
ds^{2} = G_{MN} dX^M dX^N = e^{-2\sigma} \eta_{\mu\nu} dx^\mu dx^\nu + dy^2 ,
\eeq
where $X^M = (x^\mu,y)$ denote the 5-dimensional coordinates, 
$\eta_{{\mu\nu}} = \mathit{diag}(-1,+1,+1,+1)$, $\sigma(y) = k |y|$ and 
$0\leq y \leq \LL$.  The gauge bosons are assumed to propagate in the 
bulk, while the standard model Higgs is localized on the IR brane at 
$y = \LL$. We will also assume that the fermions propagate in the bulk 
since in this case the question of proton decay can be more naturally
addressed, as we discuss below.

An important issue is how the grand unification group is broken to the
standard model group.  In the higher dimensional context there are
various possibilities.  These include dynamical symmetry breaking
through the Higgs mechanism (as in 4-dimensional models), as well as
intrinsically higher dimensional mechanisms where the gauge symmetry
is broken by nontrivial boundary conditions~\cite{orbifold}.  Orbifold
breaking seems to be a powerfull ingredient in the building of
realistic models.  When the symmetry is broken by orbifold boundary
conditions, the GUT symmetry is not an exact symmetry of the theory
since there are special points in spacetime where the symmetry is
reduced to a smaller group.  Operators living on the special points
are generally not GUT-symmetric, and can spoil the GUT predictions. 
In particular, it is necessary to make further assumptions about the
size of the localized terms that need not respect the GUT symmetry. 
One way to proceed is based on NDA, which assumes that all couplings
get strong at the cut-off scale of the effective 5-dimensional theory. 
This provides a well defined framework in which the symmetry violating
terms are sufficiently suppressed to be able to discuss unification
quantitatively.

Let us emphasize that in order to achieve unification in the
non-supersymmetric case, relevant threshold corrections at the grand
unification scale must be present.  In warped extra dimensions, these
corrections arise naturally if the gauge symmetry is broken by bulk
Higgs VEV's~\cite{Agashe:2002pr}.  As we will discuss below, for light
KK boson masses to be compatible with present phenomenological
constraints, the GUT symmetry also needs to be broken by orbifold
boundary conditions on the IR brane.  Thus, in this work, we shall
assume that the GUT symmetry is broken by a bulk Higgs VEV, as well as
by orbifold boundary conditions on the IR brane.

The dependence of the 3-2-1 gauge couplings measured by low energy
observers (at energy scales $q \ll k e^{-kL}$) on the GUT scale in
this type of compactifications has been studied in previous articles
\cite{gaugerunning,Goldberger:2002cz}, and it was found that, contrary
to what happens in flat extra dimensions, the differences between
low-energy couplings depend logarithmically on the fundamental scales
of the theory and are given by:
\beq
\label{running}
\frac{1}{g^2_i (q)} - \frac{1}{g^2_j (q)} =
\left( \frac{1}{{g^i_{UV}}^{\!\!\!\!\!\!\!2}} 
- \frac{1}{{g^j_{UV}}^{\!\!\!\!\!\!\!2}} \right)
+ \left( \frac{1}{{g^i_{IR}}^{\!\!\!\!\!2}} 
- \frac{1}{{g^j_{IR}}^{\!\!\!\!\!2}} \right)
+ \frac{(b_i-b_j)}{8\pi^2} \log\frac{k}{q}
+(\Delta_i -\Delta_j) kL
+{\cal{O}}(1)~.
\eeq
Here the first two terms can be thought of as the ``bare'' values of
the localized couplings, while the $\log$ term arises from loop
effects and is identical to the contribution that is found in purely
4-dimensional theories.  The bulk couplings get renormalized in a
universal manner and, apart from threshold corrections, cancel in the
differences.  The $\Delta_i$ terms correspond to possible threshold
corrections arising from the GUT breaking VEV's of bulk Higgs fields,
and have the important property of being enhanced by $k\,L$
\cite{Agashe:2002pr}.  Finally, the ${\cal{O}}(1)$ represents further
subdominant contributions.  The leading, GUT violating, logarithmic
effects in Eq.~(\ref{running}) can be understood as arising from the
RG evolution of the UV brane kinetic terms. It was shown in 
Ref.~\cite{Goldberger:2002cz} that, for $k \gg q \gg k\,e^{-k\LL}$,
\beq
\label{UVbraneterm}
\frac{1}{{g^i_{UV}}^{\!\!\!\!\!\!2}\;(q)}=\frac{1}{{g^i_{UV}}^{\!\!\!\!\!\!\!2}} +
\frac{b_i}{8 \pi^2} \log\frac{k}{q} + {\rm{GUT~symmetric}}~,
\eeq
which exhibits precisely the large logs found in the low-energy calculation.
The precise $\beta$-function coefficients depend on which fields
propagate in the bulk and most importantly on whether those fields
have a significant overlap with the UV brane.  In fact, in the RS
scenario the heavy KK modes of all fields are localized towards the IR
brane and therefore each KK mode gives only a tiny contribution to the
low-energy gauge coupling.  Although the sum of all of them can give a
contribution as large as the $\log$ in Eq.~(\ref{UVbraneterm}), when
including the full GUT field content the heavy KK modes give a
universal contribution that cancels in the differential ``running". 
Thus, the difference in the 3-2-1 low-energy gauge couplings is
effectively determined by the zero modes alone.  This is one way to
understand the effective four dimensional behavior.

The gauge zero modes always contribute with full strength to $b_i$
since their wavefunctions are flat.  For the fermions, however, it is
possible to add a bulk mass term, $M$, whose effect is to control the
localization of the zero mode.  When $M > k/2$, the zero mode is
localized towards the UV brane and fully contributes to the gauge
coupling running.  However, when $M < k/2$ the zero mode is
exponentially suppressed at the UV brane and effectively decouples
from the running above the TeV scale.  There is also an intermediate
region in which the fermions contribute only partially to the
renormalization of the gauge coupling.  Then, splitting the gauge and
fermion contributions, the $\beta$-function coefficients are given by
$b_1 = n_{f,1}$, $b_2= - 22/3 + n_{f,2}$ and $b_3= - 11 + n_{f,3}$,
where the $n_{f,i}$ terms are associated with the fermions.  Note that
if the fermions come in complete multiplets, then they contribute
universally: $n_{f,1} = n_{f,2} = n_{f,3}$.  However, we have seen
that considerations related to proton decay forces us to split them
into different multiplets, which can then have different bulk masses. 
In practice, we will assume universal fermion bulk masses and, as a
result, the fermions do not have an impact on whether the gauge
couplings unify or not.

It is important to emphasize that within the framework of grand
unification in warped extra dimensions, there is always a tower of
$XY$ gauge bosons with masses of order $k\,e^{-k\LL} \sim {\rm{TeV}}$,
where in general $XY$ denote the non-standard gauge bosons arising as
a remnant of the broken grand unified symmetry.  In fact, this is an
essential feature which allows one to express the differences in the
zero mode couplings, Eq.~(\ref{running}), in terms of large $\log$s of
only the four dimensional beta functions.  The key point is that the
higher KK modes always come in complete GUT multiplets (the heavy KK
modes of the broken and unbroken gauge bosons are approximately
degenerate in mass) and do not influence the coupling differences in a
large way~\cite{gaugerunning}.  However, this remarkable feature, a
clear distinction of the RS-style grand unification from the more
conventional 4-dimensional unification scenario, is also a potential
problem phenomenologically.  If the quarks and leptons are unified in
GUT multiplets, these gauge bosons will induce proton decay at an
unacceptable rate.  One way of avoiding this problem is to invoke
boundary conditions that break the GUT symmetry, which implies that
some of the components within a given multiplet will get TeV masses. 
To accommodate the SM spectrum it is then necessary to double the
number of bulk matter multiplets, with the result that quarks and
leptons reside in different multiplets.  For example, for an $SU(5)$
symmetry the zero mode $d_R$ and lepton doublet of a given family
arise from {\em different} $\bar{5}$'s.  Thus, these models do not
have a true unification of SM quarks and leptons.  On the other hand
the dangerous couplings of broken gauge bosons to pairs of zero mode
fermions are absent and cannot mediate proton decay.  One still has to
worry about higher dimension operators.  For example, if baryon
violating operators are present on the IR brane, they will only be
suppressed by the TeV scale.  Thus, it is necessary to forbid such
operators up to a very high order.  One possibility is to impose some
sort of gauged baryon symmetry, which is broken only on the UV brane
\cite{Agashe:2002pr}.  All the generated baryon violating operators
will then be suppressed by the Planck scale.

For the analysis of the electroweak constraints, we will concentrate
on an effective theory valid just above the TeV scale, and use the
previous considerations to motivate certain relations among the
effective parameters.  We can then match to an effective 4-dimensional
theory and use the well measured low-energy data to fix these
parameters.

\section{The 4-dimensional Effective Theory}

We are interested in determining the bounds on the effective
compactification scale, $k\,e^{-k\LL}$, in the class of models
described in the previous section.  These arise both from deviations
from the standard model in the zero-mode sector of the theory, as well
as from the effects mediated by the Kaluza-Klein towers of the various
bulk fields.  The former are associated with deformations of the
zero-mode wavefunctions due to the electroweak breaking Higgs VEV,
which is localized on the IR brane.  The latter arise from exchange of
KK mode gauge bosons or fermions between zero mode fields.  The gauge
KK towers induce four-fermion operators that can be important in some
regions of parameter space.  Since these are tree-level effects, they
can potentially impose the strongest constraints on the model.  In
addition, due to the large top Yukawa coupling, the top KK tower can
induce important contributions to the $T$ parameter, even though these
arise at loop level.  Similarly dangerous contributions can arise from
the KK modes of the broken gauge bosons.  The simplest way to analyze
these effects is to first obtain the effective 4-dimensional theory. 
In this section, we derive the general form of the low-energy theory
and determine a region in parameter space where the top loop effects
can be neglected.

In order to study the low-energy constraints, as well as the
properties of the lowest lying KK modes, we can consider an effective
5-dimensional theory with renormalized parameters at the matching
scale $\sim k\,e^{-k\LL}$.  The effective action describing the
standard model sector (i.e. the standard model fields together with
their corresponding KK towers) is then given by
\beqa
\label{fullgaugeaction}
S &=& \int_0^\LL d^4x dy \sqrt{-G} \left\{ - \frac{1}{2g_5^2} W_{MN}^+ W^{MN}_-
- \frac{1}{4g_5^2} W_{MN}^3 W^{MN}_3  \right. \nonumber \\
& & \hspace{2.87cm} \left. \mbox{} - \frac{1}{4{g'_5}^2} B_{MN} B^{MN} +
{\cal{L}}_\Psi - 2 \delta(y) {\cal{L}}_{UV}
- 2 \delta(y-\LL) {\cal{L}}_{IR} \right\}~,
\eeqa
with
\beqa
\label{UVbraneaction}
{\cal{L}}_{UV} &=& \frac{1}{2 g_{UV}^2} W_{\mu\nu}^+ W^{\mu\nu}_- +
\frac{1}{4 g_{UV}^2} W_{\mu\nu}^3 W^{\mu\nu}_3
+ \frac{1}{4 {g'_{UV}}^{\!\!\!\!\!\!\!2}} B_{\mu\nu} B^{\mu\nu}~, \\
\label{IRbraneaction}
{\cal{L}}_{IR} &=& \frac{1}{2 g_{IR}^2} W_{\mu\nu}^+ W^{\mu\nu}_- +
\frac{1}{4 g_{IR}^2} W_{\mu\nu}^3 W^{\mu\nu}_3
+ \frac{1}{4 {g'_{IR}}^{\!\!\!\!\!2}} B_{\mu\nu} B^{\mu\nu} \nonumber \\
& & \hspace{0cm} \mbox{} + v^2 W_\mu^+ W^\mu_-
+ \frac{1}{2} v^2 (W_\mu^3 - B_\mu)^2  + {\cal{L}}(h,\Psi)~,
\eeqa
where $W^\pm$, $W^3$ are the $SU(2)$ gauge bosons and $B$ corresponds
to the $U(1)_Y$ vector, with SM hypercharge normalization.  According
to the picture discussed in the previous section, and assuming for
example a $SU(5)$ grand unified group, the parameters in the action
Eq.~(\ref{fullgaugeaction}) obey the GUT relations $g'_5 \approx
\sqrt{\frac{3}{5}} g_5$ and $g'_{IR} \approx \sqrt{\frac{3}{5}}
g_{IR}$, while $g'_{UV}$ and $g_{UV}$ differ by the additional
logarithmic terms in Eq.~(\ref{running}).  For simplicity, we neglect
the threshold corrections $\Delta_i$ in Eq.~(\ref{running}), which
would contribute to the bulk gauge couplings.  In this way, we can
capture the leading loop effects arising from bulk fields and may
proceed with a tree-level analysis.

The SM Higgs is located on the IR brane, as required for the RS
solution to the hierarchy problem.  It has a potential which generates
a VEV $\langle H^0 \rangle \equiv \sqrt{2} v$ of order $k$, but due to
the background warping, the observable VEV appears red-shifted to $\vv
= v\,e^{-k\LL} \approx 123\,{\rm{GeV}}$.  This induces electroweak
symmetry breaking, with localized mass terms generated for the
electroweak gauge bosons.  ${\cal{L}}(h,\Psi)$ contains the Higgs
kinetic and potential terms, as well as the Yukawa couplings
\beq
\label{Yukawas}
\lambda_u \tilde{H} \bar{\Psi}_Q \Psi_{t_R} + \lambda_d H \bar{\Psi}_Q \Psi_{b_R} +
\lambda_e H \bar{\Psi}_L \Psi_{e_R}~,
\eeq
where $Q = (t_L, b_L)$ and $L = (\nu_L, e_L)$ denote the $SU(2)$ quark
and lepton doublets and $t_R$, $b_R$ and $e_R$ denote the quark and
lepton $SU(2)$ singlets.  For simplicity, we have omitted the
generation indices.  Note that all of these are 5-dimensional fermions
that, after a KK decomposition, include both left- and right-handed
towers.  The $L$ and $R$ subscripts simply indicate which one of these
towers contains a zero mode after the orbifold projection.  Also,
unlike the 4-dimensional SM, the Yukawa coupling matrices $\lambda_u$,
$\lambda_d$ and $\lambda_e$ have mass dimension $-1$.

We have not explicitly written the gluon terms, as they are irrelevant
for the electroweak constraints.  We will discuss the bulk lagrangian
for the fermions, ${\cal{L}}_\Psi$, which contains the kinetic as well
as possible bulk mass terms, below.

\subsection{Integrating out the Kaluza-Klein tower in the Gauge Sector}

We begin by integrating out the massive Kaluza-Klein modes of the
electroweak gauge sector of the theory.  It will be sufficient to do
so at tree-level.  In the presence of the gauge kinetic terms required
by unification in the Randall-Sundrum scenario, as well as the
localized Higgs VEV, $v$, it is difficult to diagonalize the quadratic
part of the gauge action exactly and, as a result, the identification
of the massive eigenstates that need to be integrated out is somewhat
tricky.  However, the electroweak precision measurements require that
$v$ be much smaller than the curvature $k$, and it becomes possible to
treat the electroweak breaking effects perturbatively.  Thus, we can
make a KK decomposition assuming that $v = 0$, integrate out these
``unperturbed" massive modes, and include the effects of $v$ in the
low-energy effective theory as a perturbation.

It will be enough to consider the relevant part of the
5-dimensional action, Eq.~(\ref{fullgaugeaction}), that describes
the $SU(2) \times U(1)$ neutral gauge bosons.  It can be written as
follows:
\beqa
\label{neutralaction}
S_N &=& \int_0^\LL d^4x dy \left\{
\frac{1}{2} W_\mu^3 {\cal{O}}^{\mu\nu} W_\nu^3  +
\frac{1}{2} B_\mu {\cal{O}}'^{\mu\nu} B_\nu +  {\cal{L}}_\psi -
2 \delta(y-\LL)\frac{1}{2} \vv^2 (W_\mu^3 - B_\mu)^2  \right\}~,
\eeqa
where $\vv = v e^{-k\LL}$,
\beq
\label{gaugeoperator}
 {\cal{O}}^{\mu\nu} \equiv \frac{1}{g_5^2} \left[ P^{\mu\nu} +
 \eta^{\mu\nu} \partial_y \left( e^{-2\sigma} \partial_y \right) +
2 \delta(y) r_{UV} P^{\mu\nu} + 2 \delta(y-\LL) r_{IR} P^{\mu\nu} \right]~,
\eeq
and $ {\cal{O}}'^{\mu\nu}$ is obtained from Eq.~(\ref{gaugeoperator})
by the replacements $g_5 \ra g'_5$, $r_{UV} \ra r'_{UV}$ and $r_{IR}
\ra r'_{IR}$.  Here we introduced the quantities $r_i \equiv
g^2_5/g^2_i$ and $r'_i \equiv {g'_5}^2/{g'_i}^2$, which have
dimensions of length.  As explained in the previous section, the
assumption of unification implies that (assuming SU(5), for example)
$g'_5 \approx \sqrt{\frac{3}{5}}g_5$ and $g'_{IR} \approx
\sqrt{\frac{3}{5}}g_{IR}$, so that $r'_{IR} \approx r_{IR}$.  The UV
brane parameters, $r_{UV}$ and $r'_{UV}$ account for the difference in
the low-energy couplings of $SU(2)$ and $U(1)$.  In
Eq.~({\ref{gaugeoperator}}) we also defined the transverse operator
$P^{\mu \nu} \equiv \eta^{\mu \nu} \partial^2 - \partial^\mu
\partial^\nu$.  For the analysis of the low-energy constraints, it is
clearly important to keep track of the fermion-gauge boson
interactions,
\beq
{\cal{L}}_\psi \supset - e^{-3\sigma} \bar{\Psi} \Gamma^\mu(T^3 W_\mu^3 +
Y B_\mu) \Psi~,
\eeq
where $\Psi$ denotes a bulk standard model fermion, $T^3$ is the third
weak iso-spin generator and $Y$ the hypercharge.

We are interested in the effective theory valid for momenta $p \ll
k\,e^{-k\LL}$, obtained by integrating out the massive KK modes at
tree-level.  As mentioned above, we can treat the electroweak symmetry
breaking VEV as a perturbation.  We could then express the action
Eq.~(\ref{neutralaction}) in terms of the ``unperturbed" KK
wavefunctions, integrate out the heavy KK states and sum up their
effects on the gauge zero modes (which will get a mass of order $g^2
\vv^2$ after including $v$ perturbatively).  It is however simpler to
treat the heavy KK modes as a single entity by working in the
5-dimensional picture and integrating out the higher dimensional
gauge field with the zero-mode subtracted.  More precisely, we write
the 5-dimensional gauge fields as
\beqa
\label{decomposition}
W_\mu^3(x,y) &=& g W_\mu^{(0)}(x) + \tilde{W}_\mu^3(x,y)  \nonumber \\
B_\mu(x,y) &=& g' B_\mu^{(0)}(x) + \tilde{B}_\mu(x,y)~,
\eeqa
and require the orthogonality conditions
\beqa
\label{orthogonality}
\int_0^\LL dy \: W_\mu^{(0)}  \tilde{W}_\nu^3 \left[ 1 + 2 \delta(y) r_{UV} +
2 \delta(y-\LL) r_{IR} \right] &=& 0 \nonumber \\
\int_0^\LL dy \: B_\mu^{(0)}  \tilde{B}_\nu \left[ 1 + 2 \delta(y) r'_{UV} +
2 \delta(y-\LL) r_{IR} \right] &=& 0~,
\eeqa
which ensure that $\tilde{W}_\mu$ and $\tilde{B}_\mu$ contain exactly
the part of the gauge fields with a nonzero ``momentum" in the extra
dimension (corresponding to the heavy modes in the KK picture).  In
order to obtain the correct normalization of the zero-modes
$W_\mu^{(0)}$ and $B_\mu^{(0)}$, we factored out explicitly the zero-th
order gauge couplings defined by
\beq
\label{0thgaugecouplings}
g^2 = \frac{g_5^2}{\LL+r_{UV}+r_{IR}}~, \hspace{1cm} 
{g'}^2 = \frac{{g'_5}^2}{\LL+r'_{UV}+r_{IR}}~.
\eeq
It is very convenient to consider the propagator associated with
$\tilde{W}_\mu$, $\tilde{B}_\mu$, which is given by
\beq
\label{KKG}
\GG_{\mu\nu} = G_{\mu\nu} - G^{(0)}_{\mu\nu}~,
\eeq
where $G_{\mu\nu}$ is the full (unperturbed) gauge propagator and
$G^{(0)}_{\mu\nu}$ the propagator for the zero-mode.  Both quantities
on the right-hand-side of Eq.~(\ref{KKG}) are easily calculated.  For
example, for $W^3_\mu$ (in $W^3_5 = 0$ gauge), the full propagator in
mixed position and momentum space has the form
\beq
\label{tensorpropagator}
G^3_{\mu\nu}(p;y,y') = \frac{P_{\mu\nu} }{p^2} G^3_p(y,y') - \frac{p_\mu p_\nu}
{p^2} G^3_0(y,y')~.
\eeq
Here $p^2 \equiv - \eta_{\mu\nu} p^\mu p^\nu$ is timelike in the 
physical region and
\beq
\label{propagator}
G^3_p(y,y') = - \frac{\pi g_{5}^{2}e^{k(y+y')}}{k(AD - BC)}
\left[ A J_1\left( \frac{p}{k} e^{k y_<} \right) -
B Y_1\left( \frac{p}{k} e^{k y_<} \right) \right]
\left[ C J_1\left( \frac{p}{k} e^{k y_>} \right) -
D Y_1\left( \frac{p}{k} e^{k y_>} \right) \right]~,
\eeq
where $J_{\alpha}$, $Y_{\alpha}$ are Bessel functions of
order $\alpha$, $y_{<(>)}$ are the smallest (largest) of $y$, $y'$ and
\beqa
\label{coefficients}
A &=& Y_0\left(\frac{p}{k}\right) +
p \, r_{UV} Y_1\left(\frac{p}{k}\right) \nonumber \\
B &=& J_0\left(\frac{p}{k}\right) +
p  \, r_{UV} J_1\left(\frac{p}{k}\right) \nonumber \\
C &=& Y_0\left(\frac{p}{k}e^{k \LL}\right) -
p  \, e^{k \LL} r_{IR} Y_1\left(\frac{p}{k}e^{k \LL}\right) \\
D &=& J_0\left(\frac{p}{k}e^{k \LL}\right) -
p  \, e^{k \LL} r_{IR} J_1\left(\frac{p}{k}e^{k \LL}\right)~. \nonumber
\eeqa
The zero-mode propagator, on the other hand is simply given by
\beq
G^{(0)}_{\mu\nu} = \frac{g^2}{p^2} \eta_{\mu\nu}~.
\eeq

Replacing Eq.~(\ref{decomposition}) in the action Eq.~(\ref{neutralaction}), we
can derive the equations of motion for $\tilde{W}^3_\mu$:
\beq
{\cal{O}}^{\mu\nu} \tilde{W}^3_\nu = 
e^{-3\sigma} \bar{\Psi} \Gamma^\mu T^3 \Psi +
2 \delta(y-\LL) \vv^2 \left[(g W^\mu_{(0)} - g' B^\mu_{(0)}) 
+ \tilde{W}^\mu_3 -
\tilde{B}^\mu \right]~.
\eeq
The classical solution to this equation can be written with the help 
of the ``KK
propagator", Eq.~(\ref{KKG}), as
\beqa
\tilde{W}^3_\mu(X) &=& \int dX' \GG^3_{\mu\nu}(X,X') \left\{ e^{-3\sigma}
\bar{\Psi} \Gamma^\nu T^3 \Psi + 2 \delta(y-\LL) \vv^2 \left[(g W^\nu_{(0)} -
g' B^\nu_{(0)}) + \tilde{W}^\nu_3 - \tilde{B}^\nu \right]\right\}(X') \nonumber \\
&\approx& \int dX' \GG^3_{\mu\nu}(X,X') \left\{ e^{-3\sigma} \bar{\Psi}
\Gamma^\nu T^3 \Psi + 2 \delta(y-\LL) \vv^2 \left[g W^\nu_{(0)} -
g' B^\nu_{(0)}\right] \right\}(X')~,
\eeqa
where $X = (x^\mu, y)$, and in the second line we kept terms up to
order $\vv^2$.  A similar expression holds for $\tilde{B}$.  The
effective theory for the gauge zero-modes is obtained by replacing
these classical solutions back in the action
Eq.~(\ref{neutralaction}).  After expanding in derivatives to get an
effective action which is local, we arrive at
\beqa
\label{neutraleffectiveaction}
S_N &=& \int d^4x \left\{ - \frac{1}{4} Z_{\mu\nu}Z^{\mu\nu} - \frac{1}{4}
F_{\mu\nu}F^{\mu\nu} - \frac{1}{2} m_Z^2 Z_\mu Z^\mu - e A_\mu J_q^\mu
\right. \nonumber \\
& & \left. \mbox{} - \frac{e}{s\,c} Z_\mu \left[ (c^2 + \vv^2 G_f^3) J_3^\mu -
(s^2 + \vv^2 G_f^B) J_Y^\mu \right] - \frac{1}{2} [G_{ff}^3 J^3_\mu J_3^\mu +
G_{ff}^B J^Y_\mu J_Y^\mu] + \cdots
\right\}~.
\eeqa
where we neglected higher derivative terms, as well as terms of order
$\vv^4$ and higher.  In the above we defined the zero-mode fermion
currents $J_3^\mu(x) = \bar{\psi}(x)\Gamma^\mu T^3 \psi(x)$,
$J_Y^\mu(x) = \bar{\psi}(x) \Gamma^\mu Y \psi(x)$ and $J_q^\mu(x) =
J_3^\mu(x) + J_Y^\mu(x)$, and used the shorthand notation
\beqa
\label{convolutions}
G_f &\equiv& \int_0^\LL dy \GG_0(\LL,y) |f^{(0)}(y)|^2 \nonumber \\
G_{ff} &\equiv& \int_0^\LL dy dy' |f^{(0)}(y)|^2 \GG_0(y,y') |f^{(0)}(y')|^2~,
\eeqa
where $f^{(0)}(y)$ is the appropriate fermion zero-mode wavefunction 
(as defined in subsection \ref{subsec:fermions}).
The propagators, $\GG_0(y,y')$, in Eqs.~(\ref{convolutions}) are now
given by Eq.~(\ref{propagator}), with the zero-mode part subtracted as
in Eq.~(\ref{KKG}) and evaluated at zero momentum, $p = 0$.  The
superscripts $3$ or $B$ in the propagator terms appearing in
Eq.~(\ref{neutraleffectiveaction}) refer to the $W^3$ and $B$
respectively.  We also wrote the action in the photon-Z basis
\beqa
A_\mu &=& s W_\mu^{(0)} + c B_\mu^{(0)} \nonumber \\
Z_\mu &=& c W_\mu^{(0)} - s B_\mu^{(0)}~,
\eeqa
where $c = g/\sqrt{g^2+g'^2}$ and $s = g'/\sqrt{g^2+g'^2}$.  Finally,
the $Z$ mass is given by
\beq
\label{Zmass}
m_Z^2 = \frac{e^2 \vv^2}{s^2 c^2} \left\{1 + \vv^2 [\GG^3_0(\LL, \LL) +
\GG^B_0(\LL, \LL)] + {\cal{O}}(v^4) \right\}~.
\eeq
The second line in Eq.~(\ref{neutraleffectiveaction}) contains the
corrections to the $Z$-boson gauge couplings to fermions due to the
deformation of the zero-mode wavefunctions induced by the Higgs VEV,
and the four-fermion interactions induced by the electrically neutral
KK towers.
The first term in the second line of Eq.~(\ref{neutraleffectiveaction})
 can be rewritten as
\beqa
-\frac{e}{s_* c_*} Z^{1/2}_{z*} [J^\mu_3 - s_*^2 J^\mu_q],
\eeqa
where
\beqa
\label{s*}
s_*^2 = s^2 \left[1 + \tilde{v}^2 \left( \frac{c^2}{s^2} G_f^B - G_f^3
\right)\right]
\eeqa
and
\beqa
\label{Z*}
Z_{z*} = 1 + \frac{\tilde{v}^2}{s^2 c^2}
\left( c^2 G_f^B + s^2 G_f^3 \right).
\eeqa
We shall use Eqs.(\ref{s*}) and Eq. (\ref{Z*}) in computing
precision electroweak parameters later on.

The charged gauge sector can be handled in a similar fashion and it is
easy to see that the result of integrating out the massive states is
\beqa
\label{chargedeffectiveaction}
S_C &=& \int d^4x \left\{ - \frac{1}{2} W^+_{\mu\nu}W_-^{\mu\nu} -
m_W^2 W^+_\mu W_-^\mu
\right. \nonumber \\
& & \left. \mbox{} - \frac{e}{\sqrt{2}s} (1 + \vv^2 G_f^3)
\left[W^+_\mu J_+^\mu + W^-_\mu J_-^\mu \right] -
\frac{1}{2} G_{ff}^3 J^+_\mu J_-^\mu + \cdots
\right\}~.
\eeqa
where 
\beq
\label{Wmass}
m_W^2 = \frac{e^2 \vv^2}{s^2} \left\{1 + \vv^2 \GG^3_0(\LL, \LL) 
+ {\cal{O}}(v^4)
\right\}~.
\eeq
and $J_\pm^\mu(x) = \bar{\psi}(x)\Gamma^\mu T_\pm \psi(x)$, with
$T_\pm = T^1 \pm i T^2$, are the charged fermion currents.

\subsection{Properties of Bulk Fermions}
\label{subsec:fermions}

We now consider the fermions in more detail.  
The properties of bulk fermions in an AdS$_5$ background
are well-known (see e.g. \cite{Gherghetta:2000qt}), but we summarize
the relevant results, both for the sake of completeness and to
establish notation.  For simplicity we shall ignore the effect of local
kinetic terms for the fermion fields. In this case,
the free action for a 5-dimensional fermion is
given by
\beqa
\label{fermionaction}
S_\Psi &=& -\int_0^\LL d^4\!x\,dy \sqrt{-G} \left\{
i \overline{\Psi}\,\Gamma^A {e_{\! A}}^{\! M} D_M \Psi +
i M \overline{\Psi} \Psi  \right\}~,
\eeqa
where $\Gamma^A$ are the flat space 5-dimensional gamma matrices,
$D_M$ is a covariant derivative, with respect to both gauge and
general coordinate transformations, $G$ is the metric defined in
Eq.~(\ref{lineelement}), ${e_{\!  A}}^{\!  M}$ is the corresponding
vielbein, and $M
= c \sigma'$ is an (odd) bulk mass term.  Expanding the fermion field
in Kaluza-Klein modes, $\Psi_{L,R}(x,y) = e^{3\sigma/2} \sum_n
\psi^{(n)}_{L,R}(x) f^n_{L,R}(y)$, one finds that the KK mode
wavefunctions so defined satisfy
\beq
\left[ \partial_y - \frac{1}{2} \sigma' \pm M \right] f_{L,R}^n =
\pm m_n e^\sigma f_{R,L}^n~,
\eeq
where the $+$ ($-$) sign applies to the wavefunction of left-handed
(right-handed) 4-dimensional fermions [we take $P_{L,R} =
\frac{1}{2} (1 \pm \Gamma^5)$].  The orthonormality conditions
required to obtain canonically normalized kinetic terms in the
effective 4-dimensional theory read
\beq
\label{normalization}
\int_0^\LL dy f^n(y) f^m(y) = \delta_{nm}~.
\eeq
Of special relevance are the zero-modes, which satisfy
\beq
\left[ \partial_y - \left(\frac{1}{2} \mp c\right) \sigma'\right] f_{L,R}^0 = 0~,
\eeq
and are therefore given by
\beq
f_{L,R}^0(y) = \sqrt{\frac{k(1\mp 2c)}{e^{(1\mp 2c) k \LL}-1}}
e^{(\frac{1}{2} \mp c) \sigma}~.
\eeq
Depending on the fermion $Z_2$ parity, one of these zero-modes is
projected out and the remaining one has an exponential profile that is
localized towards one of the orbifold fixed points (depending on
whether $c > 1/2$ or $c < 1/2$).  In order to simplify the discussion
we will adopt a convention for the bulk mass term of a given fermion
$f$ such that $c_f > 1/2$ ($c_f < 1/2$) corresponds to the physical
zero mode being localized towards the UV brane (IR brane).  For
example, if $\Psi_Q$ and $\Psi_u$ denote the 5-dimensional fermion
fields corresponding to the left-handed quark doublet and right-handed
up quark $SU(2)$ singlet respectively, then we define $c_Q$ and $c_u$
by
\beqa
\label{Qufermionaction}
S &=& -\int_0^\LL d^4\!x\,dy \sqrt{-G} \left\{
i \overline{\Psi}_Q\,\Gamma^A {e_{\! A}}^{\! M} D_M \Psi_Q +
i \overline{\Psi}_u\,\Gamma^A {e_{\! A}}^{\! M} D_M \Psi_u \right.\nonumber \\
& & \hspace{3.2cm} \left. \mbox{} + i c_Q \sigma' \overline{\Psi}_Q \Psi_Q -
i c_u \sigma' \overline{\Psi}_u \Psi_u + \cdots \right\}~,
\eeqa
Therefore, the zero-mode wavefunctions will always be given by
\beq
\label{zeromodefermion}
f^0(y) = \sqrt{\frac{k(1 - 2 c_f)}{e^{(1 - 2 c_f) k \LL}-1}}
e^{(\frac{1}{2} - c_f) \sigma}~.
\eeq

We will also need the higher KK mode wavefunctions.  With the above
convention for the bulk mass term, the $Z_2$ even and odd solutions
for a given 5-dimensional fermion field, denoted by $f^n_+(y)$ and
$f^n_-(y)$ respectively, are
\beqa
f^n_\pm(y) &=& A_n e^{\sigma} \left[ J_{|c_f \pm \frac{1}{2}|}
\left( \frac{m_n}{k} e^{\sigma} \right) + b\left( m_n/k \right)
Y_{|c_f \pm \frac{1}{2}|} \left( \frac{m_n}{k} e^{\sigma} \right) \right]~,
\eeqa
where
\beq
b(x) = - \frac{J_{|c_f - \frac{1}{2}|}(x)}{Y_{|c_f - \frac{1}{2}|}(x)}~.
\eeq
The KK masses are determined by the condition $b(m_n/k) = b(e^{k\LL}
m_n/k)$ and the normalization constants $A_n$ are determined by
Eq.~(\ref{normalization}).

The localized Higgs VEV, which is responsible for the zero-mode
fermion masses, induces mixing between the previous zero-mode and the
massive KK tower in a way which is analogous to the gauge sector
described in the previous section.  This mixing is proportional to
both the bare 5-dimensional Yukawa coupling and to the values of
the KK wavefunctions at the VEV position, $y = \LL$.  The zero-mode
wavefunctions at $y = \LL$ can be obtained directly from
Eq.~(\ref{zeromodefermion}).  As for the massive KK modes, we find
that, for the lowest lying ones and for all moderate $c_f$, the
wavefunctions evaluated at the IR brane are given to a good approximation by
\beq
\label{fnL}
f^n_+(\LL) \approx \pm \sqrt{2 k}~.
\eeq

These mixing effects can be important in the top sector.  In fact,
since the effective 4-dimensional top Yukawa coupling is close to one,
it is necessary to move the top zero-mode wavefunctions closer to the
IR brane ($c_f < 1/2$), or otherwise one would obtain an exponential
suppression of the effective 4-dimensional Yukawa coupling.  This
would require one to invoke a 5-dimensional Yukawa coupling which is
non-perturbative.  Even for values of $c_f$ which result in
perturbative Yukawa interactions, the strong mixing enhances the loop
effects of the KK modes of the top, which can render the theory
incompatible with electroweak precision measurements.

The Yukawa term that couples the even towers associated 
with the left- and right-handed top has the following structure
\beqa
\label{topYukawa}
\sqrt{-G} \delta(\LL-y) \frac{\hat{\lambda}_5}{\Lambda}
\sqrt{2} v \, \overline{\Psi}_{t_L} \Psi_{t_R}
&\approx& \delta(y-\LL) m_t \left[ \overline{\psi}^{(0)}_{t_L} \psi^{(0)}_{t_R} +
\frac{\sqrt{2k\LL}}{a_{t_R}} \sum_{n \neq 0} \overline{\psi}^{(0)}_{t_L}
\psi^{(n)}_{t_R} \right.
\nonumber \\ & & \hspace{1cm} \left. \mbox{} +
\frac{\sqrt{2k\LL}}{a_{t_L}} \sum_{n \neq 0} \overline{\psi}^{(n)}_{t_L}
\psi^{(0)}_{t_R} +
\frac{2k\LL}{a_{t_L} a_{t_R}} \sum_{n,m \neq 0} \overline{\psi}^{(n)}_{t_L}
\psi^{(m)}_{t_R} \right]
\eeqa
where we wrote the 5-dimensional top Yukawa coupling in
Eq.~(\ref{Yukawas}) in units of the cutoff scale $\Lambda$ and
introduced the dimensionless coupling $\hat{\lambda}_5$.  We also
defined the parameters
\beq
\label{af}
a_{f} = \sqrt{\frac{(1 - 2 c_{f}) k \LL}{e^{(1-2c_{f})k \LL}-1}}\,
e^{(1/2-c_f)k\LL} \approx
\left \{
\begin{array}{ll}
\sqrt{(2 c_{f} - 1) k \LL}\,e^{-(c_f-1/2)k\LL} & \vspace{2mm}
\hspace{0.5cm} \;\; c_f - 1/2 \gsim 1/2k\LL \\
1 & \hspace{0.5cm}\;\; c_f = 1/2 \\
\sqrt{(1-2 c_{f}) k \LL} & \hspace{0.5cm}\;\; 1/2 - c_f \gsim 1/2k\LL
\end{array}
\right.
\eeq
and the zero-mode mass
\beq
m_t = a_{t_L} a_{t_R} \frac{\hat{\lambda}_5}{\Lambda \LL}\sqrt{2} \vv~.
\eeq

First let us see what is required to reproduce the observed top mass,
i.e. $y_t \equiv a_{t_L} a_{t_R} \hat{\lambda}_5/(\Lambda \LL) \sim
1$.  The volume suppression factor $1/(\Lambda \LL)$ indicates that we
may want to make $\hat{\lambda}_5$ as large as possible.  A similar
situation arises in the gauge sector.  Since the 4-dimensional
(unified) gauge coupling $g_4^2 \approx \hat{g}_5^2/(\Lambda \LL)$ is
of order one, and $\Lambda \LL > k \LL \sim 30$, we are led to take
the dimensionless constant $\hat{g}_5$ as large as possible.  Thus, it
is natural to assume that both these interactions become strong at the
cutoff scale $\Lambda > k$ and we can use NDA to estimate the size of
the dimensionless couplings $\hat{g}_5$ and $\hat{\lambda}_5$.  In
this strong coupling limit, we have \cite{Manohar:1983md}
\beq
\hat{g}_5 \sim l_5~, \hspace{1.5cm} \hat{\lambda}_5 \sim
\frac{l_5}{\sqrt{l_4}}~,
\eeq
where $l_5 = 24 \pi^3$ and $l_4 = 16 \pi^2$ are the five- and
4-dimensional loop factors respectively.  Requiring that $g_4 \sim
1$, we find $\Lambda \LL \sim l_5$.  Note that this gives $\Lambda/k
\sim l_5/(k\LL) \sim 25$ which shows that the gauge interactions get
strong at a scale where the theory looks 5-dimensional, and
therefore justifies the use of the above 5-dimensional NDA
analysis.

The top Yukawa coupling is then
\beq
y_t \sim \frac{a_{t_L} a_{t_R}}{\sqrt{l_4}} = \frac{a_{t_L} a_{t_R}}{4 \pi}~,
\eeq
which, according to Eq.~(\ref{af}), can easily be of order one if
$c_{t_L}$, $c_{t_R}$ are slightly below $1/2$.

As an aside, we may ask whether the gravitational interactions also
get strong at the scale $\Lambda$ defined above.  This will be the
case if the 5-dimensional Planck scale, $M_5$, is related to
$\Lambda$ by $M_5^3 \sim \Lambda^3/l_5$.  However, if we take $k \sim
10^{16}\,{\rm{GeV}}$ as suggested by gauge coupling unification and
$M_P = 2 \times 10^{18}\,{\rm{GeV}}$, we find
\beq
\frac{M_5^3}{\Lambda^3/l_5} \approx \frac{k M_P^2}{\Lambda^3/l_5} =
l_5 \left(\frac{M_P}{k}\right)^2 \left(\frac{k}{\Lambda}\right)^3 \sim 10^3~.
\eeq
Therefore, for the previous choice of parameters the gravitational
interactions are still weak when the gauge (and top Yukawa)
interactions get strong.

Now let us go back to the effects of the massive KK top towers and the
interactions given in Eq.~(\ref{topYukawa}).  Since the requirement
that the top mass be reproduced led us to consider $c_{t_L}, c_{t_R} <
1/2$, it follows from Eqs.~(\ref{topYukawa}) and (\ref{af}) that the
mixing effects are proportional to $m_t \sqrt{2k \LL} /a_{t_{L,R}}
\approx m_t/(\frac{1}{2} - c_{t_{L,R}})^{1/2}$.  These couplings can
be considerably large if $c_f$ is close to $1/2$ (but still $1/2 - c_f
> 1/2k\LL$), thus imposing some further constraints on the allowed
region of parameter space.  In fact, even if these factors are small
enough that we can treat these effects in the mass insertion
approximation, there can still be sizable contributions to the $T$
parameter due to loops of the top KK modes.  We estimate these
contributions in the next section.

\section{Precision Electroweak Analysis}

As described above, there are two main contributions affecting the
precision electroweak observables in this theory.  The first type are
the Higgs localization effects that are associated with a deformation
of the zero-mode weak gauge boson wavefunctions and masses.  These
effects are related to the functions $G_f$ and $\GG_0(L,L)$ introduced
in the previous section.  When all fermions have a common value of the
bulk mass parameter $c_f$, these affect the couplings of the gauge
bosons to quarks and leptons in a universal way.  They are therefore
associated with oblique corrections.  Also in this class are
loop-level corrections to the gauge boson self-energies due to the KK
modes of bulk fields.  The second type of contributions arise from
exchange of the KK modes of the gauge bosons at tree-level, leading to
non-oblique corrections proportional to the functions $G_{ff}^{3,B}$
introduced in the previous section.  Since the Higgs localization
effects are associated with oblique corrections, their impact on all
experimental observables can be described in terms of parameters $S$,
$T$ and $U$, defined in Ref.~\cite{Peskin:1991sw}.  These parameters
provide an excellent description of all precision electroweak data in
the case that the non-oblique corrections, parametrized by $G^i_{ff}$,
are much smaller than the oblique ones, namely $G^i_{ff} \ll G^i_0, \;
G^i_f$.

Whenever the non-oblique corrections are non-negligible, the above
procedure cannot be applied.  However, there are certain cases in
which effective parameters $S_{eff}$, $T_{eff}$ and $U_{eff}$ may be
defined in order to describe only a subset of the experimental data,
which, due to its precision, leads to the most stringent experimental
tests on the theory.  This is the case, for instance, whenever the
heavy KK modes of the neutral gauge fields are sufficiently heavy, the
only impact of non-oblique corrections on $m_W$ and the Z-pole
observables comes indirectly from the Fermi constant $G_F$.  Using
$G_F$ as an input value to the precision electroweak data, the
non-oblique corrections induced by the heavy KK modes data can be
easily absorbed into the definition of new, effective parameters
$T_{eff}$ and $U_{eff}$~\cite{Carena:2002dz}.  The advantage of this
procedure is that the functional dependence of $m_W$ and the $Z$-pole
observables on the new effective parameters is the same as the
functional dependence of these observables on $S$, $T$ and $U$ in the
case in which only oblique corrections are present, and thus, one can
use the fits of $S$, $T$, and $U$ derived from such observables to
understand the physical picture and constraint the parameters of the
model.

\subsection{Massive Kaluza-Klein effects}

We start by considering the effects of the heavy KK modes in the
theory.  First, due to the large top Yukawa coupling, the top KK modes
can induce potentially large contributions to the $T$ parameter.  In
fact, we saw in Eq.~(\ref{topYukawa}) that the localized Higgs VEV
induces large mixings between the top-quark left and right zero modes
and their even KK towers.  Furthermore, the heavy KK modes receive
electroweak breaking contributions to their masses that are enhanced
by factors of $2 k \LL$.\footnote{These factors arise from the KK mode
wavefunctions at the IR brane, and are also responsible for the strong
coupling of the KK modes to IR brane fields.} The largest
contributions to $T$ are dominated by graphs with the heavy KK modes
running in the loop.  These may be estimated in the mass insertion
approximation.  The contribution induced by the mixing of the first KK
modes of the left- and right-handed top quark is of order
\beq
\label{deltaTtop}
\Delta T_{t^1_L,t^1_R} \sim \left(\frac{2k\LL}{a_{t_L} a_{t_R}}\right)^4
\left(\frac{m_t}{m^{(1)}_{t}}\right)^2 \left[ \frac{N_c}{16 \pi s^2 c^2}
\left(\frac{m_t}{m_Z}\right)^2\right]~,
\eeq
where $m^{(1)}_{t}$ is the mass of the heaviest of these two KK modes,
$a_{t_L}$ and $a_{t_R}$ were defined in Eq.~(\ref{af}), $s$ is the
sine of the weak mixing angle, and the term in square brackets is the
SM top contribution, which is of order one.  Note that, due to the
brane localized Yukawa interactions and the consequent mixing among KK
modes exhibited in Eq.~(\ref{topYukawa}), summing up the KK tower
leads to a quadratic sensitivity on the cutoff $\tilde{\Lambda} =
\Lambda\,e^{-k \LL}$.\footnote{This is similar to the model of
Ref.~\cite{Barbieri:2000vh}, which contained large, uncalculabe,
contributions to the $\rho$ parameter.  Note, however, that in our
case the compactification scale is about an order of magnitude
higher.} This quadratic sensitivity depends strongly on the assumption
that the couplings of the heavy KK modes are well described by
Eq.~(\ref{topYukawa}).
For example, the presence of IR brane localized kinetic terms for the
top and bottom quark fields can suppress the couplings between the Higgs
and the higher top KK modes \cite{delAguila:2003kd} and lead to a
milder sensitivity to the cutoff scale.  Since we expect that the
contribution of the lightest KK mode remains approximately unaffected
by these effects, Eq.~(\ref{deltaTtop}) provides a lower bound on the
total effect of integrating out the heavy KK modes.

An upper bound on the order of magnitude of the contribution of the KK
tower on $T$ may be obtained by a naive extrapolation of the sum over
all KK modes up to the cutoff scale, which at the end of
subsection~\ref{subsec:fermions} was estimated as $\tilde{\Lambda}
\lsim 25 k\,e^{-k \LL}$.
For $c_{t_L},c_{t_R} < 1/2$ it reads $\Delta T \lsim
60~(\frac{1}{2}-c_{t_L})^{-2} (\frac{1}{2}-c_{t_R})^{-2}
(m_t/m^{(1)}_{t})^2$.  Therefore, this contribution may be strongly
suppressed for values of $c_f \ll 1/2$.

However, as shown in Ref.~\cite{Carena:2002dz}, when the fermions are
localized towards the IR brane, $c_f \ll 1/2$, there are important
non-oblique corrections and, in the absence of local gauge kinetic
terms the bounds on the KK scale become tight~\cite{Carena:2002dz}. 
One possibility is to take the bulk mass parameters of the first and
second generation larger than the one associated with the third
generation.  However, choosing different bulk masses for the fermions
can potentially lead to large FCNC effects~\cite{Carone:1999nz}.  In
particular, taking the third generation $c_Q$ very different from the
first two generation $c_f$'s can lead to dangerous contributions to $Z
\to b \bar{b}$~\cite{Hewett:2002fe}.  Thus, in order to keep
non-oblique corrections small, while suppressing the dangerous
contributions to the $Z \to b \bar{b}$ decay width and flavor changing
neutral currents we are led to take equal $c_f$'s for the left-handed
top and the first two generations of fermions.  The constraints on
$t_R$ are much weaker and it is possible to render the above
contributions to $T$ negligible by taking $c_{t_R}$ sufficiently
negative.

There are also contributions that are sensitive only to $c_{t_L}$,
which are induced by the $\psi_{t_L}^n$--$\psi_{t_R}^0$ mixing.  The
effect of the first KK level is given by
\beq
\label{deltaTtop2}
\Delta T_{t^1_L,t^0_R} \sim \frac{(2k\LL)^2}{3 a_{t_L}^4}
\left(\frac{m_t}{m^{(1)}_{t_L}}\right)^2 \left[ \frac{N_c}{16 \pi s^2 c^2}
\left(\frac{m_t}{m_Z}\right)^2\right]~,
\eeq
and, as before, one can estimate the KK contribution to $T$, which
scales like $\log\tilde{\Lambda}$, by summing up the effects of the KK
modes lighter than $\tilde{\Lambda}$.  For $\Lambda \sim 25 k$, these
type of contributions give $\Delta T_{KK} \lsim
4~(\frac{1}{2}-c_{t_L})^{-2}$~$(m_t/m^{(1)}_{t_L})^2$.

The KK tower effects computed above can be considered as a
conservative estimate of the total contribution to the $T$ parameter. 
Such effects can be rendered small by taking appropriate values of
$c_{t_L}$ and $c_{t_R}$.  For instance, considering $c_{t_L} \simeq
0.3$ and $c_{t_R} \simeq -5$, and using $m^{(1)}_{t_L} \gsim 2.5 k
e^{-k \LL}$, these contributions are smaller than about 1/10 of the
dominant contribution induced by the oblique corrections to the gauge
boson masses and wave functions, that will be discussed in more detail
in the next section.  Therefore, for $c_f = c_{t_L} \simeq 0.3$,
$c_{t_R} \simeq -5$, where $c_f$ stands for the rest of the fermions,
the above corrections have a negligible impact on the fit to the
experimental data.

Let us remark that the effects of the physics above $\tilde{\Lambda}$
can be parametrized in the effective theory by local operators such as
\begin{equation}
{\cal{O}}_T \; = \; \delta(y-\LL) \; \frac{c}{\Lambda^2} \;
\left( H^\dagger D_\mu H \right)^2~.  
\end{equation}
Even when the unknown coefficient attains its strong coupling value,
$c \sim 16 \pi^2$, their contribution to $T$ is negligible for
$\Lambda \sim 25 k$ and $k\,e^{-k\LL}$ above a few TeV.

There are possible additional effects induced by the presence of the
GUT theory.  The KK modes of the heavy $X$ and $Y$ bosons of the
theory are essential for the question of unification of couplings, but
can also induce important contributions to the precision electroweak
observables.  This is due to the fact that these gauge bosons form a
doublet under $SU(2)$ and, if coupled to the Higgs, can lead to large
contributions to the $T$ parameter, because the splitting in masses
squared will be of order $\vv^2$, while their masses are of order TeV.
In fact, if these gauge bosons are even about the IR brane, so that
the IR brane respects the GUT symmetry, they give a contribution to
$T$ of order
\beq
\label{deltaTXY}
\Delta T_{XY} \sim \frac{N_c (2k\LL)^2}{4 \pi s^2} \left(\frac{\vv}
{m^{(1)}_{XY}}\right)^2~,
\eeq
which is of order one when $m^{(1)}_{XY}$, the mass of the first KK
mode of the gauge bosons $X$ and $Y$ is of order of a few TeV. A
straightforward way of cancelling these effects is by demanding the
$X$ and $Y$ gauge bosons to be odd under the $Z_2$ orbifold associated
with the IR brane.  In this case, their couplings to the Higgs vanish
and these additional contributions to the electroweak precision
observables become negligible, because the $XY$ masses are highly
degenerate.

\subsection{Effective S, T and U parameters}

Having established that the possible effects from the heavy KK modes
of the fermions and the GUT sector of the theory can be made
negligible, we turn to the effects associated with the zero-mode
sector and the non-oblique contributions to $G_F$.  {}From the
expression of the effective action for the charged gauge currents and
$m_W$, Eqs.~(\ref{chargedeffectiveaction}) and (\ref{Wmass}),
respectively, one finds that $G_F$ is given by
\beqa
\label{G_F}
4 \sqrt{2} G_F = \frac{1}{\tilde{v}^2} \left[ 1 + \tilde{v}^2 \left( 
2 G_f^{3} -  \GG_0^{3}(L,L) - G_{ff}^{3} \right) \right].
\eeqa
The last term in the above equation represents the non-oblique
corrections. Observe that, with this normalization 
$\tilde{v} \simeq 123$~GeV.

In the case the non-oblique corrections, proportional to the
four-fermion interactions $G_{ff}$ are negligible, the corrections to
all experimental data at the Z-pole, can be taken into account via the
standard $S$, $T$ and $U$ parametrization.  Following the procedure
outlined in Ref.~\cite{Peskin:1991sw} and using Eqs.~(\ref{s*}),
(\ref{Z*}) and (\ref{Wmass}), one find the values
\beqa
\label{STU}
\alpha S &\approx& 4 \vv^2 [s^2 G_f^3 + c^2 G_f^B] + {\cal{O}}(\vv^4)~, \nonumber \\
\alpha T &\approx& \vv^2  [2 G_f^B - \GG_0^B(\LL,\LL)]
+ {\cal{O}}(\vv^4)~,  \\
\alpha U &\approx& {\cal{O}}(\vv^4)~, \nonumber
\eeqa

As mentioned above, in many cases the only relevant non-oblique
corrections to $m_W$ and the Z-pole observables come indirectly
through the Fermi constant $G_F$.  In this case, following
Ref.~\cite{Carena:2002dz}, and considering the expression of $G_F$,
Eq.~(\ref{G_F}), it is possible to define the effective parameters
$S_{eff}$, $T_{eff}$ and $U_{eff}$, which are given by,
\beqa
\label{STUeffective}
\alpha S_{\rm{eff}} &\approx& 4 \vv^2 
[s^2 G_f^3 + c^2 G_f^B] + {\cal{O}}(\vv^4)~, \nonumber \\
\alpha T_{\rm{eff}} &\approx& \vv^2  [2 G_f^B - \GG_0^B(\LL,\LL)
+ G^3_{ff}] + {\cal{O}}(\vv^4)~,  \\
\alpha U_{\rm{eff}} &\approx& 
- 4 s^2 \vv^2 G^3_{ff} + {\cal{O}}(\vv^4)~, \nonumber
\eeqa

For completeness, we shall also give the expression for $\rho_*(0)$,
which is defined as the low-energy ratio of neutral to charged current
interactions.  These low-energy observables depend on the following
combinations
\beq
\Delta^{3,B} = 2 G_f^{3,B} - G_{ff}^{3,B} - \GG_0^{3,B}(\LL,\LL)~.
\label{Delta}
\eeq
In terms of $\Delta^{3,B}$, the effective charged four-fermion
interaction, which determines the Fermi constant, $G_F$, from muon
decay, is given by
\beq
{\rm{CC}} = \frac{1}{\vv^2} 
\left\{1 + \vv^2 \Delta^3 + {\cal{O}}(\vv^4) \right\}~\equiv G_F.
\eeq
Similarly, the neutral four-fermion interaction that is relevant in
neutrino-nucleon scattering can be written as
\beq
{\rm{NC}} = T^3_\nu \left(T^3_q - \bar{s}^2 Q_q\right) \frac{1}{\vv^2}
\left\{1 + \vv^2 [\Delta^3 + \Delta^B ] 
+ {\cal{O}}(\vv^4) \right\}~,
\eeq
where
\beq
\bar{s}^2 = s^2 \left\{1 + \vv^2 \left[ G_{ff}^3 -G_f^3 -
\frac{c^2}{s^2} (G_{ff}^B - G_f^B)\right]  + {\cal{O}}(\vv^4) \right\}~.
\eeq
It follows that
\beq
\rho_*(0) = 1 + \vv^2 \Delta^B + {\cal{O}}(\vv^4)~.
\eeq

It is important to stress that, in the case in which all SM fermions
are localized on the infrared brane, $G_{ff}^i = G_f^i = G_0^i(L,L)$
and therefore $\Delta^{3,B} = 0$.  Consequently, in this case
$\rho_*(0) = 1$, while $\bar{s}^2 = s^2$.

\subsection{Explicit Expressions for $G_f$, $G_{ff}$ and $\GG_0(\LL,\LL)$}
\label{gfgffg0}

We have established that the most important new effects in the
low-energy theory are contained in $G_f$, $G_{ff}$ and
$\GG_0(\LL,\LL)$, which we now evaluate in terms of the fundamental
parameters of the 5-dimensional theory.  For simplicity, in this
subsection we assume that we have a simple gauge group.  The
application to the $SU(2) \times U(1)$ theory is straightforward: one
simply plugs the relevant expressions for the $SU(2)$ and $U(1)$
sectors into the formulae of the previous section.  As we have seen in
Eqs.~(\ref{Zmass}) and (\ref{Wmass}), the corrections to the gauge
boson masses are determined by $ \GG_0(\LL, \LL)$, which is given by
\beq
\label{GLL}
 \GG_0(\LL, \LL) = - \frac{e^{2k\LL} g^2 }{k^2}
\frac{2 k^2 (\LL + r_{UV})^2 - 2 k (\LL + r_{UV}) + 1}
{4 k (\LL + r_{UV} + r_{IR})} ~.
\eeq

The corrections to the fermion-gauge boson vertices and the effects of
the massive KK modes are contained in $G_f$ and $G_{ff}$.  These
depend on the fermion zero-mode wavefunctions and therefore on the
parameter $c_f$, which we are taking to be common for all the fermions
with the exception of $c_{t_{R}}$, as discussed in
subsection~\ref{subsec:fermions}.  Using the zero-mode wavefunctions
given in Eq.~(\ref{zeromodefermion}) to evaluate
Eqs.~(\ref{convolutions}), their exact analytic expressions can be
obtained in a straightforward manner, but unfortunately the results
have a somewhat complicated dependence on $c_f$.  However, there are
some limits in which the expressions simplify considerably.  There are
three qualitatively different cases depending on whether $c_f -
\frac{1}{2} > 1/2k\LL$, $\frac{1}{2} - c_f > 1/2k\LL$ or $c_f \approx
\frac{1}{2}$.  These cases correspond to whether the fermion zero-mode
wavefunction is localized towards the UV brane ($c_f> \frac{1}{2}$),
towards the IR brane ($c_f < \frac{1}{2}$) or the conformal case $c_f
= \frac{1}{2}$.

$\bullet$ $c_f - \frac{1}{2} > 1/2k\LL$: The zero-mode fermions are
localized towards the UV brane.  Neglecting exponentially small terms
we find that,
\beqa
\label{GfUVfermions}
G_f &=& \frac{e^{2k\LL} g^2 }{k^2} \frac{k(\LL + r_{UV}) - 1 - k r_{IR} +
2 k^2 r_{IR} (\LL + r_{UV})}{4 k (\LL + r_{UV} + r_{IR})}~, \\
\label{GffUVfermions}
G_{ff} &=& - \frac{e^{2k\LL} g^2 }{k^2} \frac{2 k^2 r_{IR}^2 + 2 k r_{IR} +
1}{4 k (\LL + r_{UV} + r_{IR})} ~,
\eeqa
where $g$ is the (zero-th order) zero-mode gauge coupling defined in
Eq.~(\ref{0thgaugecouplings}) and we have dropped exponentially
suppressed terms of order $e^{-kL}$.  The results in
Eqs.~(\ref{GfUVfermions}) and (\ref{GffUVfermions}) are independent
of $c_f$.

{}From these expressions we also find that the $\Delta$ parameter,
Eq.~(\ref{Delta}), that
enters in the low-energy observables is given by
\beq
\Delta = 2 G_f -  \GG_0(\LL, \LL) - G_{ff} = \frac{e^{2k\LL} g^2 }{2k^2} k
(\LL+r_{UV}+r_{IR})~.
\eeq

$\bullet$ $c_f = \frac{1}{2}$: This is the conformal case in which the
zero-mode fermions couple with equal strength at all points along the
extra dimension.  We find
\beqa
\label{GfFlatfermions}
G_f &=& \frac{e^{2k\LL} g^2 }{k} \frac{r_{UV}(1+2 k^2 L r_{IR}) -
k r_{UV}(\LL + r_{UV} + r_{IR}) + r_{IR} (2 k^2 \LL^2-2 k \LL+1)
}{4 k^2 \LL (\LL + r_{UV} + r_{IR})}~,  \\
\label{GffFlatfermions}
G_{ff} &=& - e^{2k\LL} g^2 \frac{r_{UV}^2  - 2 (k \LL -1) r_{UV} r_{IR} +
(2 k^2 \LL^2 - 2 k \LL +1) r_{IR}^2}{4 k^3 \LL^2 (\LL + r_{UV} + r_{IR})}~,
\eeqa
and
\beq
\Delta = \frac{e^{2k\LL} g^2 }{2 k^2}
\left( \frac{2 k^2 \LL^2 - 2 k \LL + 1}{2 k^2 \LL^2} \right)
k (\LL+r_{UV}+r_{IR})~.
\eeq
Note that Eqs.~(\ref{GfFlatfermions}) and (\ref{GffFlatfermions})
vanish when $r_{UV} = r_{IR} = 0$.  This is a consequence of the gauge
orthogonality condition, Eq.~(\ref{orthogonality}), and the fact that
in the conformal case the fermion zero-mode wavefunction is flat and
therefore proportional to the gauge zero mode wavefunction.  Thus, the
coupling of the zero-mode fermions to the higher KK gauge modes
vanishes identically in this case.

$\bullet$ $\frac{1}{2} - c_f > 1/2k\LL$: When the fermions are
localized towards the IR brane the expressions for $G_f$ and $G_{ff}$
have a very complicated dependence on $c_f$ and we do not present them
here.  However, the linear combination corresponding to $\Delta$
simplifies to
\beq
\Delta = \frac{e^{2k\LL} g^2 }{4 (1 - c_f) k^2} k (\LL+r_{UV}+r_{IR})~.
\eeq

Finally, we emphasize again that in the limit where the fermions are
localized on the IR brane, we have
\beq
\label{IRfermions}
G_f = G_{ff} = \GG_0(\LL, \LL) ~,
\eeq
where $ \GG_0(\LL, \LL)$ was given in Eq.~(\ref{GLL}).  These
expressions can also be formally obtained by taking the limit $c_f \ra
- \infty$.  

Note that, after multiplying by $\vv^2$, all the corrections are of
order $g^2v^2/k^2$ times a function of $k\LL$, $k r_{UV}$ and $k
r_{IR}$, which may represent an enhancement or a suppression depending
on the location of the fermions.  The corrections to the gauge boson
masses, determined by $\GG_0(\LL,\LL)$, are always enhanced by a
factor of order $k \LL$.  When the fermions are localized towards the
UV brane, we have $G_{ff} \ll G_f \ll \GG_0(\LL,\LL)$.  The conformal
case is somewhat special, since $G_{f}$ and $G_{ff}$ vanish when
$r_{IR} = r_{UV} = 0$.  For the values of the brane kinetic
coefficients that interest us~\footnote{The values of the local brane
couplings $r_{UV}$ have been estimated by assuming, as in
Eq.~(\ref{UVbraneterm}), that they include all the large logarithmic
contributions, that lead to the difference between the high- and
low-energy gauge couplings}, $k r_{UV}^\prime \simeq 10$ and $k r_{UV}
\simeq - 10$, it is still true that $\GG_0(\LL,\LL)$ dominates over
$G_{f}$ and $G_{ff}$.  In the limit that the fermions are localized
towards the IR brane, one generically has $G_{ff} \lsim G_f \lsim
\GG_0(\LL,\LL)$.  Note also that unless $c_f$ is very large and
negative, $\Delta$ has always the same order of magnitude,
independently of the location of the fermions, and is enhanced by
${\cal{O}}(k \LL)$.  The behavior of $G_f$ and $G_{ff}$ as a function
of $c_f$ can be seen in Fig.~1.  In Fig.  2, we show the relative size
of $G_f$ and $G_{ff}$ compared to $\GG_0(\LL, \LL)$ for $0< c_f < 1$. 
We observe that the non-oblique corrections coming from $G_{ff}$ are
much smaller than the oblique contributions $G_f$ and $\GG_0(\LL,
\LL)$ in this region.

\begin{figure}[t]
\vspace*{-1.cm}
\centerline{ \hspace*{-0.5cm}
\epsfxsize=8.5cm\epsfysize=5.5cm
		     \epsfbox{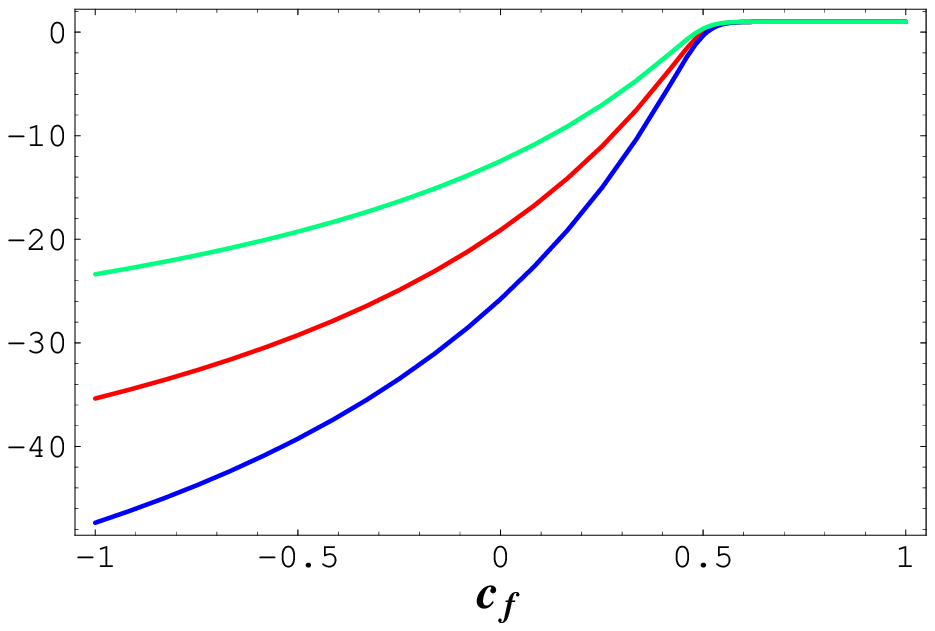} 
\epsfxsize=8.5cm\epsfysize=5.5cm
		     \epsfbox{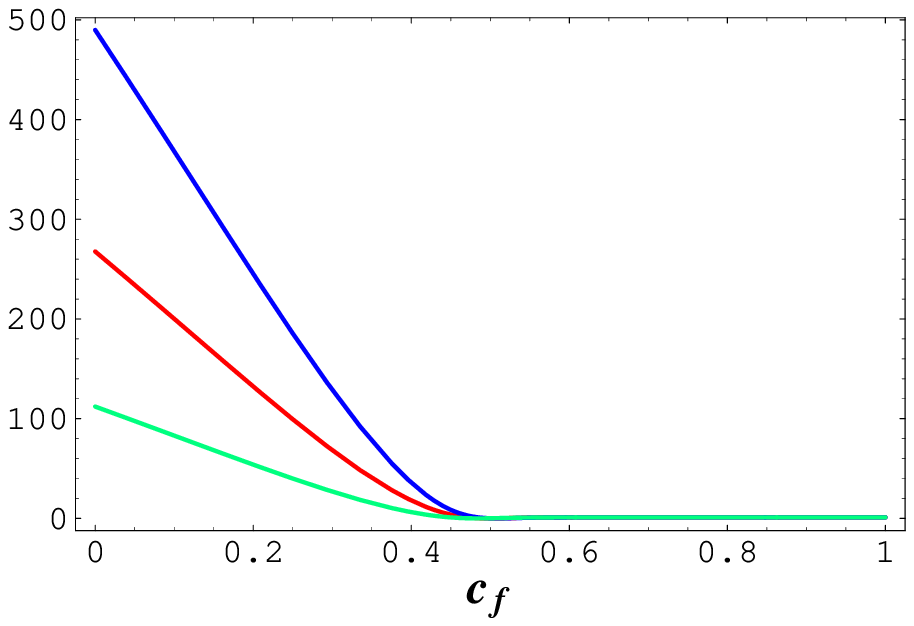}
}
\caption{Behavior of $G_f/G_f^\infty$ (left) and $G_{ff}/G_{ff}^\infty$ (right)
as a function of $c_f$, where the superscript $\infty$ indicates
$c_f = +\infty$ (fermions localized on the UV brane). The curves correspond to
$k\,r_{UV} = 0$ (red), $k\,r_{UV} = 10$ (blue) and $k\,r_{UV} = -10$ (green).
In all of them we took $k\,r_{IR} = 0$ and
$k\LL = 30$.}
\label{fig:Gf}
\end{figure}

\begin{figure}[t]
\vspace*{0.cm}
\centerline{ \hspace*{-0.5cm}
\epsfxsize=8.5cm\epsfysize=5.5cm
		     \epsfbox{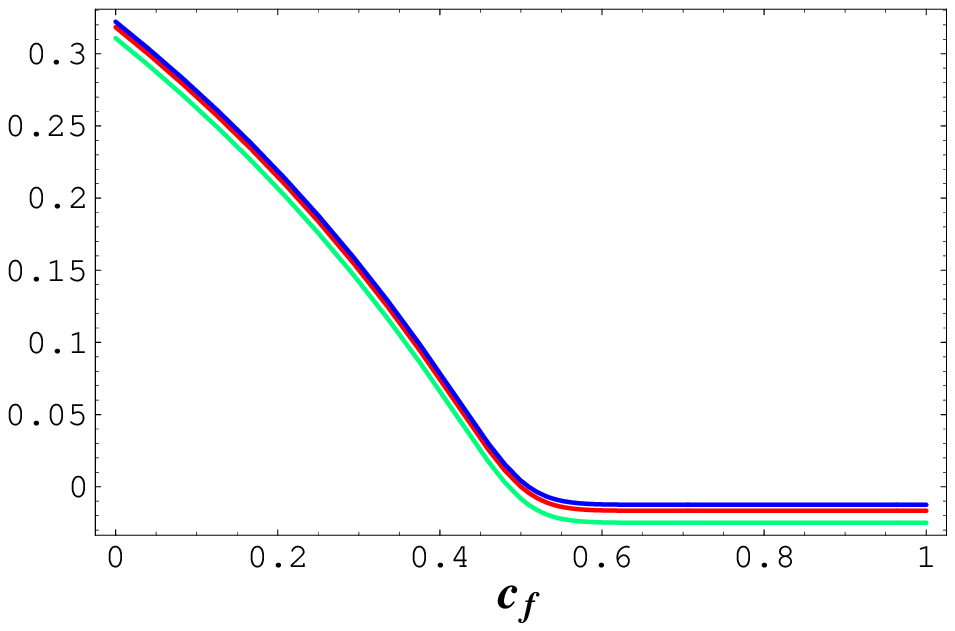} 
\epsfxsize=8.5cm\epsfysize=5.5cm
		     \epsfbox{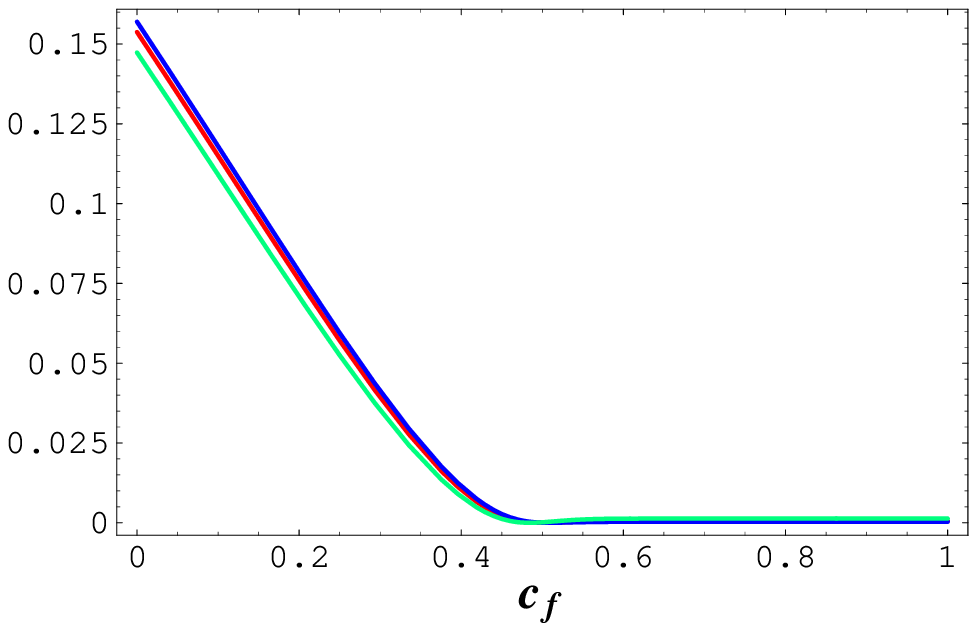}
}
\caption{Behavior of $G_f/\GG_0(\LL, \LL)$ (left) and
$G_{ff}/\GG_0(\LL, \LL)$ (right) as a function of $c_f$.
The curves correspond to $k\,r_{UV} = 0$ (red), $k\,r_{UV} = 10$ (blue)
and $k\,r_{UV} = -10$ (green). In all of them we took $k\,r_{IR} = 0$
and $k\LL = 30$.}
\label{fig:GfoverG0}
\end{figure}

\section{Precision Electroweak Measurement Constraints}

In this section, we present the analysis of the precision electroweak
data.  As we have discussed in the previous section, whenever the
effect of the four-fermion interactions $G_{ff}$ becomes negligible, the
oblique parameters $S$, $T$ and $U$, Eq.~(\ref{STU}) lead to a good
parametrization of the corrections to all relevant experimental data. 
In this case, the effective parameter $U$ is equal to zero within the
order of approximation we are using and therefore one can effectively
reduce the fit to the experimental data to a two parameter fit.  We
shall consider as input parameters the experimental values of the mass
$M_Z$, the Fermi constant $G_F$ and the electromagnetic gauge
coupling.  This will determine some of the underlying parameters of
the model.  Based on the grand unified RS scenario we take $g_5 =
\sqrt {5/3} g'_5$.  {}From Eq.~(\ref{UVbraneterm}), taking $k \sim
M_{GUT} \sim 10^{16}~\rm{GeV}$, we estimate that the size of the log
term in $r^\prime_{UV}$ should be roughly $L/3$ provided the bare
$r^\prime_{UV}$ is small and $b_1$ not much different from its SM
value.  Our results are rather insensitive to this choice.  From here,
for each given value of $L$ and $r_{IR}$, we fix $\tilde{v}$, $g_5$
and $r_{UV}$ using the experimental values of $G_F$, $\alpha$ and
$M_Z$ (c.f. Eq.  (\ref{G_F}), Eq.~(\ref{Zmass}) and
Eqs.~(\ref{0thgaugecouplings})).  We will vary the remaining free
parameters, $r_{IR}$, $L$, and the Higgs mass, to see which regions of
parameters are in accord with precision measurements.

An inspection of the results presented in section~\ref{gfgffg0} leads
to the conclusion that, whenever $c_f \gsim 0.3$, the four-fermion
interactions may be safely neglected.  On the other hand, values of
$c_{t_L}$ and $c_{t_R}$ smaller than about 0.5 may be needed in order
to avoid large contributions to the $T$ parameter induced by KK
fermion loops.  In order to minimize the contribution of the KK
fermion modes to the $T$ parameter, one can proceed with the
localization of the third family in the infrared brane, as suggested
in Ref.~\cite{Hewett:2002fe}.  Alternatively, choosing the value of
$c_{t_R} <-5$, while keeping all other fermion bulk mass parameters
$c_f \lsim 0.5$, is sufficient to suppress the most important
loop-induced contributions to the precision electroweak parameters.

In this work, we shall choose $c_{t_R} \simeq -5$, while for all other
fermions $c_f = 0.3$.  This procedure keeps all light fermions at the
same location in the bulk.  This choice of bulk mass
parameters $c_f$ has the advantage, compared to the case in which all
the third generation fermions are localized on the IR brane, that
flavor changing neutral current contributions are suppressed, with the
largest FCNC effects in the right-handed up-type sector, inducing a
$Z$-$t_R$-$c_R$ interaction\footnote{The size of this interaction can
be estimated assuming that the right-handed rotation matrices are of the order
of the corresponding CKM elements, resulting in a $Z$-$t$-$c$ coupling
strength of $10^{-6} g$.  This is well below existing low energy
bounds and is too small to affect single top production at the LHC
\cite{Tait:2000sh}.}.  Furthermore, large additional contributions to
$R_b$ arising from the mismatch of the $Z$ coupling to bottom quarks
compared to other fermions are avoided.  Finally, for these values of
the parameters, the description of the $Z$-pole observables and $m_W$
in terms of the oblique parameters $S$, $T$ and $U$ remains valid, and
should be computed by universally taking $c_f = 0.3$.  For $c_f = 0.3$
the value of $\GG_0(L,L)$ is much larger than $G_f$ and $G_{ff}$. 
From Eq.~(\ref{STU}), and the results presented in
Figure~\ref{fig:GfoverG0}, one immediately observes that the theory
tends to induce relatively large values of the $T$ parameter and small
contributions to the parameter $S$.

These extra dimensional contributions to $S$ and $T$ must be combined
with the more standard contributions from the Higgs.  Compared to some
reference Higgs mass $m^2_{ref}$ (which must be specified when
performing the fit), the Higgs contributions are \cite{Peskin:1991sw},
\bea
S_H & \simeq & \frac{1}{12 \pi}
\log \left( \frac{m_h^2}{ {m^2_{ref}}} \right) \\
T_H & \simeq & -\frac{3}{16 \pi c_0^2}
\log \left( \frac{m_h^2}{ {m^2_{ref}}} \right) \\
U_H & \simeq & 0 .
\eea
Thus, we observe that the large positive corrections to the $T$
parameter from the extra dimensions may be cancelled by the large
negative corrections associated with a massive Higgs boson in the
standard 'conspiracy' scenario familliar when there are weak scale
vector-like quarks which mix with top \cite{Dobrescu:1997nm}. 
Therefore, it is natural to expect that in this case the bounds on the
KK gauge boson masses may be relaxed by taking large values of $m_H$. 
On the other hand, a large Higgs boson mass also induces positive
corrections to the parameter $S$.  However, for $c_f \simeq 0.3$, this
contribution is also cancelled by the one coming from KK modes, which
turns out to be negative and comparable to the one coming from the
Higgs, for approximately the same values of the KK masses as the ones
needed to cancel the $T$ contribution.

\subsection{Numerical Results}

\begin{figure}[t]
\vspace*{-1.cm}
\centerline{ \hspace*{1cm}
\epsfxsize=10.0cm\epsfysize=10.0cm
		     \epsfbox{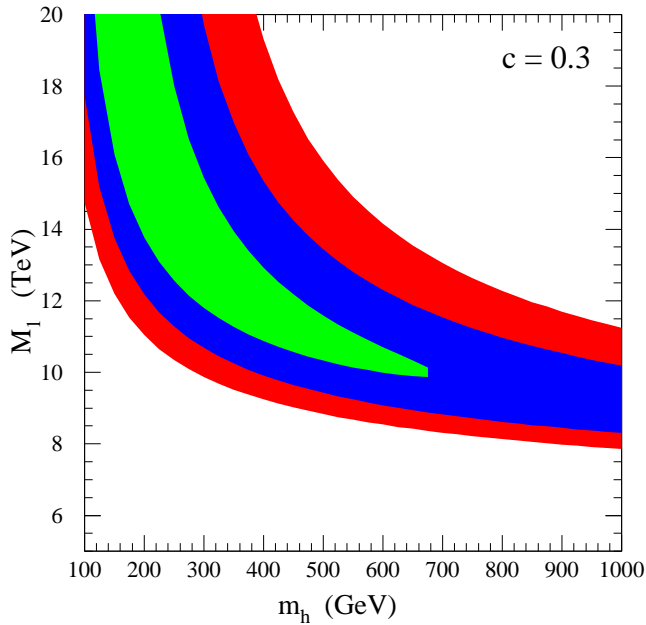}
}
\caption{Allowed bands of the RS unified model in the parameter space of
the Higgs mass and the first KK mode mass.  The central green band represents
$1\sigma$ agreement with the electroweak fit to $S$ and $T$ (with $U=0$),
whereas the surrounding blue and red bands indicate $2\sigma$ and $3\sigma$
agreement, respectively.}
\label{fig:massesc}
\end{figure}

\begin{figure}[t]
\vspace*{-1.cm}
\centerline{ \hspace*{1cm}
\epsfxsize=10.0cm\epsfysize=10.0cm
		     \epsfbox{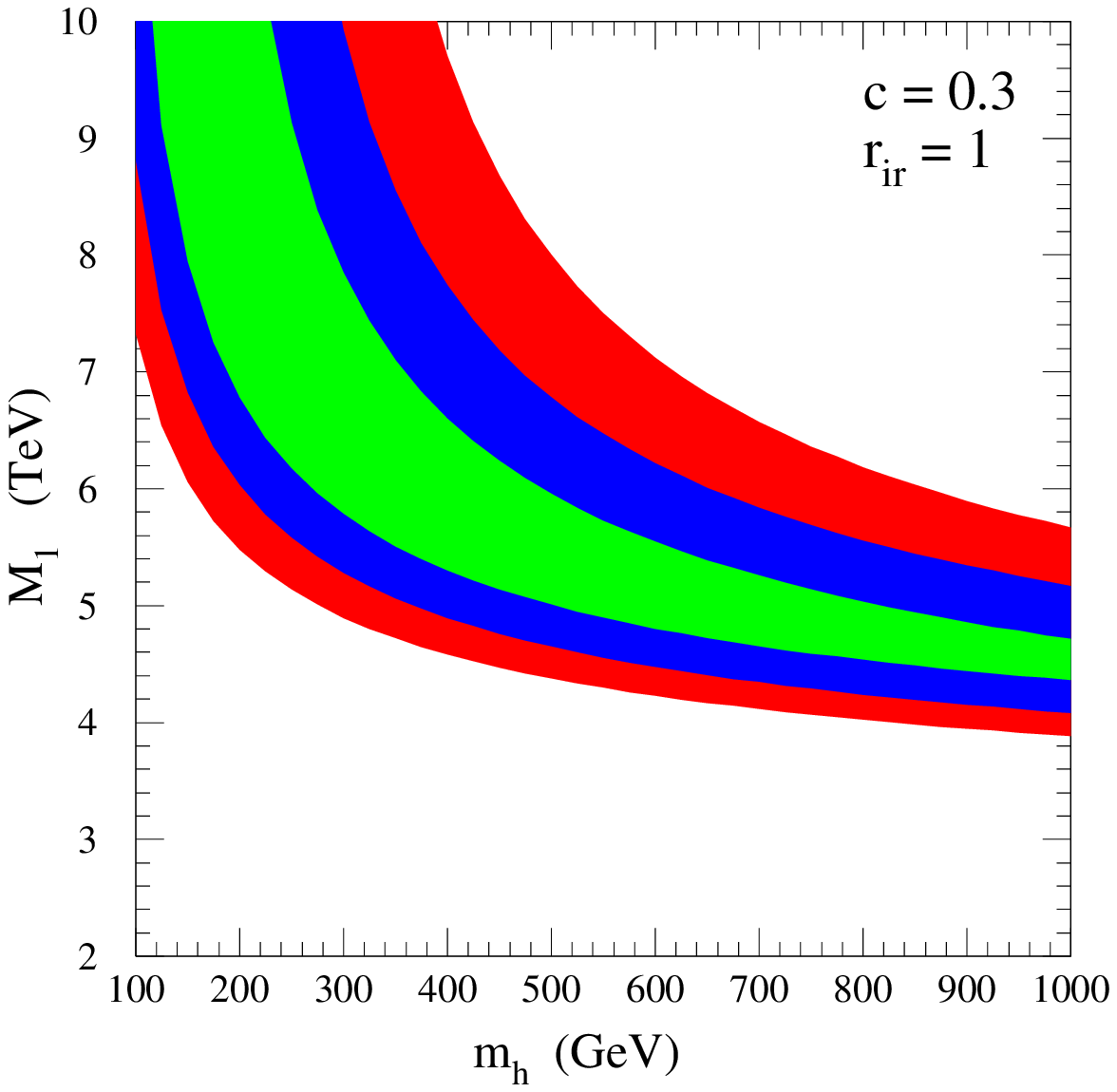} \hspace*{-1.5cm}
\epsfxsize=10.0cm\epsfysize=10.0cm
		     \epsfbox{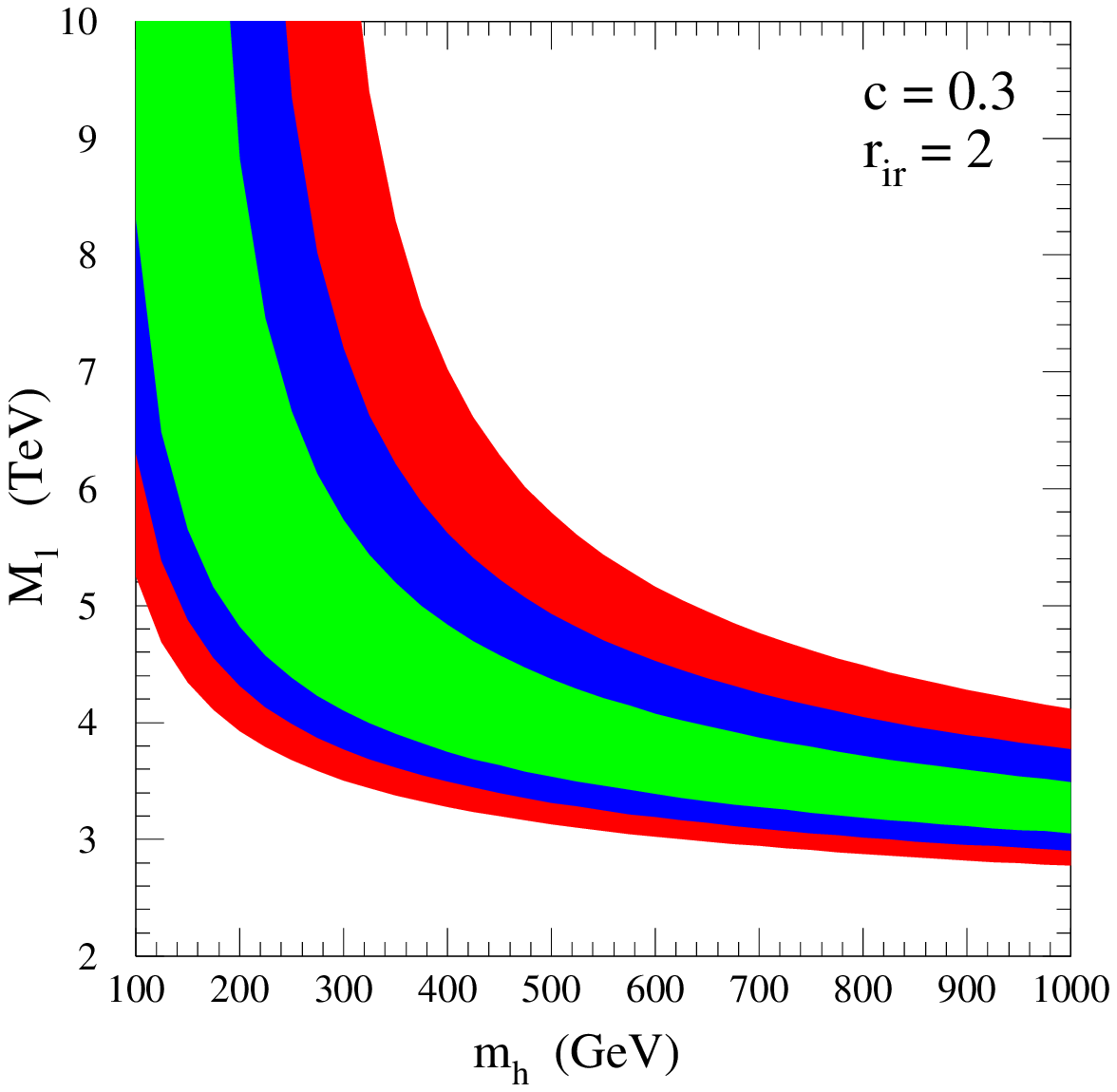}
}
\caption{As Fig.~\protect{\ref{fig:massesc}}, but with non-zero 
(unified) kinetic terms on the IR brane of $k r_{IR} = 1,2$.}
\label{fig:massesrir}
\end{figure}

In Ref.~\cite{Carena:2002dz} we demonstrated that, in the case of
light KK modes, the fit to the precision electroweak data is greatly
improved by the presence of brane gauge kinetic terms.  The main
reason for this improvement is associated with the fact that for the
same values of $k$ and $L$, the mass of the first gauge boson KK mode
becomes significantly lower for large values of $r_{IR}^i$, than in
the case of vanishing brane kinetic terms.  However, since we are
interested in the question of unification of couplings, we cannot set
$r_{IR}^i$ to arbitrarily large values.  Of course, if the $r_{IR}^i$
are unified, they will not disrupt the unification of the gauge
couplings.  However, since we must impose orbifold breaking of the GUT
symmetry on the IR brane to suppress the contributions from loops of
the broken gauge bosons to $T$, there is no reason to expect that the
various $r_{IR}^i$ would be unified, and thus the prediction of
unified gauge couplings could be lost.  Therefore, we shall only
concentrate on small values of $k r_{IR}^i \lsim 2$, assuming that
they are the same for both electroweak gauge fields, and thus do not
disrupt unification.  Note that even should they not be unified, $k
r_{IR}$ of order 2 will change the unification conditions by roughly
the same order of magnitude as the threshold corrections needed for RS
unification anyway.  Thus, our results may be taken as indicitive of
the picture for small but non-unified IR brane terms as well.

Our framework has $U \simeq 0$, and thus we choose to compare the RS
model with the LEP electroweak working group fit to $S$ and $T$ which
imposes $U=0$ in the fit \cite{EWWG}.  In
Figures~\ref{fig:massesc},\ref{fig:massesrir} we present the results
for the upper bound on the first weak gauge boson KK mass $M_{KK}$ as
a function of the Higgs mass for $c_f = 0.3$ (for all fermions except
the top right, as explained above), and for (small) different values
of $r_{IR}$.  As anticipated, the fit to the data is significantly
improved by large values of the Higgs mass $m_h \gsim 300$
GeV\footnote{Note that while such large Higgs masses are generally
incompatible with unification because the Higgs self-interaction
typically reaches a Landau pole before the GUT scale, in the RS
scenario fields localized on the IR brane see an effective cut-off of
order $\tilde{\Lambda} = \Lambda\,e^{-k\LL}$, relaxing this
constraint.}.  However, the bound on the first weak gauge boson KK
mode mass is still about 11 TeV and therefore difficult to detect at
the LHC. However, even the addition of small kinetic terms in the
infrared brane may have dramatic effects in the spectrum.  Indeed, in
the case of $k r_{IR}' = k r_{IR}^{2} = 2$ ($k r_{IR}' = k r_{IR}^{2}
= 1$) a bound of about 4~TeV (5~TeV) may be obtained for values of the
Higgs mass larger than 400 GeV. Even in the case of a light Higgs
boson, $m_h < 200 $ GeV, one can accomodate values of the gauge boson
KK masses of about 5 TeV in a way consistent with precision
electroweak data.  Observe that even for large values of the Higgs
mass, of about 1~TeV, a good fit to the precision electroweak data can
be obtained, due to the simultaneous cancellation of the $T$ and $S$
contributions coming from the Higgs and extra dimensional effects
discussed at the end of the last section.

Given the above conclusions, it seems possible that the LHC can study
the prospect of RS unification.  The electroweak data requires the
masses of the first KK gauge bosons to be around 11 TeV (a few TeV if
the $k r_{IR}$ are allowed to be as large as 2 and the Higgs mass
$\gsim 300$ GeV).  Given that this over-all scale is large compared to
$\tilde{v}$, one thus expects quasi-degenerate $SU(3)$, $SU(2)$,
$U(1)$, and $XY$ gauge bosons.  Even at 8 TeV, it may be possible to
see signs of the KK gluons indirectly, e.g. as $\bar{q}q \bar{q}q$
operators one might search for as a sign of quark compositeness.  For
masses in the range of a few TeV, there is the hope that the GUT
sector could be produced and studied, providing clear experimental
evidence of an RS unified theory.

\section{Conclusions}

Warped extra dimensional scenarios have the remarkable property of
leading to a logarithmic dependence on the fundamental scale $k$ in
the difference between the low energy couplings of the model.  Since
the logarithmic dependence is controlled by the 4-dimensional beta
functions of the theory, this implies that the question of unification
may be studied at a similar level of precision as in the four
dimensional case.

In this work we studied a particular realization of these scenarios
that is consistent with unification of couplings, and leads to the SM
as the low energy effective theory.  While the Higgs field is
localized in the IR brane, in a way consistent with the
Randall-Sundrum solution to the hierarchy problem, the gauge bosons
and the fermions fields propagate in the bulk.

We introduce a formalism that allows to study the main corrections to
the precision electroweak data in warped extra dimensions in the case
in which the gauge fields propagate in the bulk.  These corrections
are parametrized by three set of functions $G_{ff}$, $G_{f}$ and
$G_0$, associated with effects induced by the heavy KK modes as well
as the gauge boson zero mode wavefunction and mass corrections
respectively.  These functions also depend on the localization of the
fermions in the bulk.

In the case of bulk fermions which couple to the Higgs in a relevant
way, there may be other important loop corrections to the $T$
parameter.  These originate from the breakdown of the custodial
symmetry associated with the large top-quark Yukawa coupling and
constrain the right-handed top-quark mass parameter to be negative,
$c_{t_R} < 0$, while the value of the left-handed top-quark mass
parameter must be smaller than the conformal case value, $c_{Q^3_L} <
0.5$.  In our work we have chosen values of the fermion mass
parameters $c_f$ such that the contributions to flavor changing
neutral currents and to $R_b = \Gamma(Z \to b \bar{b})/\Gamma(Z \to
{\rm hadrons})$ are suppressed, while the fermion induced
contributions to the precision electroweak observables become
subdominant.

The result of our analysis shows that, with small but non-vanishing
local brane kinetic terms for the gauge fields, one can obtain a model
consistent with unfication of couplings and with precision electroweak
data with gauge boson KK masses of the order of a few TeV. The bound
on the gauge boson KK masses is correlated with the value of the Higgs
mass, and a light KK spectrum demands Higgs masses larger than 300
GeV. Thus, the RS unified framework provides a scenario which is a
novel alternative to 4-dimensional SUSY grand unification, with
potentially interesting experimental signatures at the LHC.

~\\
{\Large \bf Acknowledgements}\\
~\\
The authors are pleased to acknowledge conversations with K. Agashe
and R. Sundrum.  Work at ANL is supported in part by the US DOE, Div.\
of HEP, Contract W-31-109-ENG-38.  Fermilab is operated by
Universities Research Association Inc.  under contract no. 
DE-AC02-76CH02000 with the DOE. A.~D.~ is supported by NSF Grants
P420D3620414350 and P420D3620434350 and also wants to thank the Theory
Division of Fermilab for the kind invitation.

\end{document}